\documentclass[fontsize=12pt,paper=a4,pagesize,captions=tableheading,numbers=noenddot,toc=flat]{article}
\usepackage[dvips]{color}
\usepackage{epsfig}
\usepackage{amsmath}
\usepackage{cite}
\usepackage{graphicx}
\usepackage{color}
\usepackage{subfigure}
\usepackage{lipsum}
\usepackage{inputenc}
\usepackage[colorlinks=true]{hyperref}
\usepackage{cleveref}
\usepackage[frozencache=true,cachedir=minted-cache]{minted} 
\ifdefined\directlua
\usepackage{fontspec}
\else
\usepackage[T1]{fontenc}
\usepackage[nomath]{lmodern}
\fi

\begin{document}
\begin{center}
\Large{\bf The effect of the WGC condition on the maximal energy extracted from black holes}\\
\small \vspace{1cm} {\bf E. Naghd Mezerji$^{\star}$\footnote {Email:~~~e.n.mezerji@stu.umz.ac.ir}},
{\bf J. Sadeghi$^{\star}$\footnote {Email:~~~pouriya@ipm.ir}},
{\bf B. Pourhassan$^{\dagger}$\footnote {Email:~~~b.pourhassan@du.ac.ir}},
\\
\vspace{0.5cm}$^{\star}${Department of Physics, Faculty of Basic Sciences, University of Mazandaran\\
	 P. O. Box 47416-95447, Babolsar, Iran}\\
\vspace{0.5cm}$^{\dagger}${School of Physics, Damghan University, Damghan, 3671641167, Iran.}\\
\vspace{0.5cm}$^{\dagger}${Canadian Quantum Research Center 204-3002 32 Ave Vernon, BC V1T 2L7 Canada.}\\
\small \vspace{1cm}
\end{center}
\begin{abstract}
In this paper, we study the Penrose process and the maximum energy extracted from the collision of two particles near the Kerr-Newman black hole with WGC condition. We consider the collision process when two particles collide in the ergosphere region of a black hole, scattering two new particles; one of them falls into the black hole and the other escapes to infinity. Our calculations also show an increasing in the energy. We also checked how much energy could be received from a black hole for particles with different spins. Results of this paper will help us to identify and study the black holes of astrophysics and the particles with different spins. Also, it states how black holes work in WGC conditions.
\end{abstract}
\newpage
\tableofcontents
\section{Introduction}\label{I}
Black holes are important objects in theoretical physics. Black hole thermodynamics is the best way to study them \cite{01,02}. We can increase our knowledge about the quantum gravity by studying the perturbative \cite{03,04,05}, and non-perturbative \cite{06,07,08,09,10} quantum corrections on the black hole thermodynamics. One of the most important issues in physics is finding the consistency of the effective field theory with quantum gravity, known as Landscape. While many other effective field theories are inconsistent, they are called Swampland. The most important universal test to distinguish these two groups is weak gravity conjecture (WGC). It means that gravity is always the weakest force and shows the extremality state of the black hole, (for example $M=Q$ in Reissner-Nordstr$\ddot{o}$m black hole). We know the mass $m$, and charge $q$ of the black hole comes out of it during Hawking radiation, $M-m\geq Q-q$, and also the electric force of the initial charge is stronger than the gravitational attraction between them, $m\leq q$. As a result, if $M=Q$ is established, we have $q\geq m$. This means that the conjecture of weak gravity is permanently established.We have a little understanding of WGC and the extremal black holes of astronomy. Due to being important rotating black holes and the existence of many astrophysical black holes in the extremal condition, we examine the extremal Kerr-Newman black hole in this article. The WGC condition in the charged rotating black holes equals $M_2=a_2+Q_2$. To avoid naked singularity, we examine the near-extremality condition, $M^2=\lim_{\varepsilon\longrightarrow 0}(a^2+Q^2+\varepsilon)$.\cite{11,12,13,14,15,16,17,18,19,20,21,22,23,24,25,26,27,28,29,31,32}\\
On the other hand, Banados, Silk, and West (BSW) showed that rotating black holes (Kerr) acted like particle accelerators in 2009. They showed that if the particle’s angular momentum is double its energy ($J/M=2E/M$), the high center-of-mass energies will be calculated. This mechanism has been investigated for many black holes \cite{33,34,35,36,37,38}. We use the general ansatz of stationary axisymmetric metrics that contain different black holes to solve (\ref{e.11}). Penrose also suggested that it was possible to extract energy from the rotating black hole in ergoregion. This process is so that the particles are broken down into two particles upon entering the region. For this reason, the particles with negative energy could exist in this area;  the negative ones fall into the black hole, and the positive ones escape to the infinite. One of the most important general relativity and astrophysics processes is the Penrose process. We assume two particles collided and produced two new scattering ones (due to this process having better efficiency). One of the scattered particles with positive energy escapes from the black hole ($E_3$). We calculate the energy of this particle and show that it is equal to the energy extracted from the black holes.\cite{39,40,41,42,43,44,45,46,47,48,49,50,51,52,53,54,55,56,57,58,59,60,61,62,63,64,65,66,67,68,69}\\ 
This paper aims to consider the collided two massive spinning particles in ergoregion and obtain the maximum energy extracted from them. Also, we obtain the maximum energy at spins $s_1(= s_3)=-1.526,-1.519$ and $s_2(=s_4)=-1.513,-1.526$ with $E_3=9\times 10^5m_3$. Also, we calculate the maximum energy for all particles with different spins.\\
Article structure: In Sec.\ref{II}, we describe the BSW mechanism as comprehensive and brief. In Sec.\ref{III}, we illustrate the Penrose process. In Sec.\ref{IV}, we study the particle collision process near Kerr-Newman black holes. Finally, we have reconsideration the conclusions in Sec.\ref{V}.
\section{The BSW mechanism}\label{II}
We first briefly describe the universal mechanism of BSW; then, in the following sections, we will use these equations to calculate the variables of different black holes. We consider the general ansatz of a stationary axisymmetric black hole.
\begin{equation}\label{e.1}
ds^2=-\alpha^2(r,\theta)f(r,\theta)dt^2+\frac{\alpha^2(r,\theta)dr^2}{f(r,\theta)}+\frac{d\theta^2}{f_{\theta}(r,\theta)}+\beta^2(r,\theta)(d\phi-h(r,\theta)dt)^2
\end{equation}
We consider the space-time of parity reflection symmetric $\theta\longrightarrow\pi-\theta$, so we are only working on the equatorial plane set at $\theta=\pi/2$. The black hole horizon is in $r=r_h$, and we have $f\arrowvert_{r_h}= 0$; also the angular velocity at horizon $h(r_h)$ derive in relation $J/M=h(r,\theta)E/M$. We write the metric in tetrad form for convenience in calculations.
\begin{equation}\label{e.2}
\begin{split}
g_{ab}&=\eta_{\alpha\beta}e_{a}^{\alpha}e_{b}^{\beta}\\
e_{a}^{0}=\alpha\sqrt{f}(dt)_a,\hspace{6pt}e_{a}^{1}=\dfrac{\alpha(dr)_a}{\sqrt{f}}&,\hspace{6pt}e_{a}^{2}=\dfrac{(d\theta)_a}{\sqrt{f_\theta}},\hspace{6pt} e_{a}^{3}=\beta((d\phi)_a-h(dt)_a)
\end{split}
\end{equation}
In the BSW mechanism, spinning particles' orbits deviate from a geodesic. So, their motion is described by Mathisson-Papapetrou-Dixon (MPD) equations.\cite{25,26,27,28}
\begin{equation}\label{e.3}
\frac{DP^a}{D\tau}=-\frac{1}{2}{R^{a}}_{bcd}v^{b}S^{cd},\hspace{30pt} \frac{DS^{ab}}{D\tau}=P^{a}v^{b}-P^{b}v^{a},
\end{equation}
Here $\dfrac{D}{D\tau}$, $v^{a}=(\frac{\partial}{\partial r})^{a}$, $S^{ab}=m{\varepsilon^{ab}}_{cd}u^cs^d$, and $P^{a}=mu^a$ are covariant derivative, tangent vector, the spin tensor and 4-momentum, respectively. Also, we define the mass as $m=-P^aP_a$, and we have the condition of normalization $u^av_a=-1$. By placing in MPD, equations are the relationship between them.
\begin{equation}\label{e.4}
v^a-u^a=\frac{S^{ab}R_{bcde}u^{c}S^{de}}{2(m^{2}+\frac{1}{4}R_{bcde}S^{bc}S^{de})}.
\end{equation}
In the presence of spin, the particle momentum is not parallel to the world-line. The motion of spinning particles in the equatorial plane causes $v^2=0$, so according to the above relation, we have $u^2=0$. As a result, the only non-zero vector of spin is $s^2=-s$. Its direction is parallel to the black hole spin ($s>0$) and in the opposite direction ($s<0$). The only non-zero components in spin tensor are,
\begin{equation}\label{e.5}
S^{01}=-msu^3,\hspace{15pt}S^{03}=msu^1,\hspace{15pt}S^{13}=msu^0
\end{equation}
We obtain the conserved quantities by the Killing vectors $\chi^{a}=(\frac{\partial}{\partial t})^{a}$ and $\psi^{a}=(\frac{\partial}{\partial \phi})^{a}$.
\begin{equation}\label{e.6}
\dfrac{E}{m}=-u^{a}\chi_{a}+\frac{S^{ab}\nabla_{b}\chi_{a}}{2m},\hspace{30pt}\dfrac{J}{m}=u^{a}\psi_{a}-\frac{S^{ab}\nabla_{b}\psi_{a}}{2m}.
\end{equation}
Energy and angular momentum are two constants that the Killing vector gives us, which are rewritten as follows by functions in metrics.
\begin{equation}\label{e.7}
\begin{split}
\dfrac{E}{m}=&(2sh\beta'+s\beta h'+2\alpha^2)\frac{\sqrt{f}}{2\alpha}u^0+(h\beta+\frac{sf\alpha'}{\alpha}+\frac{sf'}{2}+\frac{s\beta^2hh'}{2\alpha^2})u^3\\
\dfrac{J}{m}=&\frac{s\beta'\sqrt{f}}{\alpha}u^0+(\frac{s\alpha f'}{2}-\frac{\beta}{h}+\frac{s\beta^2h'}{2h^2(\alpha h-1)})u^3.
\end{split}
\end{equation}
According to the normalization condition $u^au_a=(u^3)^2+(u^1)^2-(u^0)^2=-1$, (\ref{e.4}), (\ref{e.7}) and (\ref{e.8}) we can obtain
\begin{equation}\label{e.8}
\begin{split}
u^0=&\frac{A}{\sqrt{f}B},\hspace{75pt}u^1=\sqrt{\frac{A^2}{fB^2}-\frac{C^2}{B^2}-1},\hspace{19pt}u^3=\frac{C}{B},\\
v^0=&(A+D)u^0+Gu^3,\hspace{15pt}v^1=Au^1,\hspace{92pt}v^3=Hu^3-Gu^0,
\end{split}
\end{equation}
and,
\begin{equation}\label{e.9}
\begin{split}
A=&2\frac{J}{m}s\alpha^2(2f\alpha'+\alpha f')+2\beta\alpha(\frac{E}{m}-\frac{J}{m}h)(\beta sh'-2\alpha^2),\\
B=&\beta^3s^2h'^2+2\alpha^2s^2\beta'f'+4\alpha fs^2\alpha'\beta'-4\alpha^4\beta,\\
C=&-2\alpha^2(\frac{J}{m}s\beta h'+2\alpha^2\frac{J}{m}-2s\beta'(\frac{E}{m}-\frac{J}{m}h)),\\
D=&4f\alpha s^2(\alpha\beta''-2\alpha'\beta'),\\
G=&-2\sqrt{f}\beta s^2(h'(3\alpha\beta'-2\beta\alpha')+\alpha\beta h''),\\
H=&2\alpha^2\beta s^2f''-4\alpha^4\beta-3\beta^2s^2h'^2+4\beta\alpha s^2\alpha'f'+\beta f(4\alpha s^2\alpha''-4s^2\beta\alpha'^2).
\end{split}
\end{equation}
For spinning particles, we have to apply the time-like condition $v^av_a<0$ because $v^av_a$ is not a conserved quantity, leading to causality problems and superluminal motions. According to (\ref{e.9}) and the time-like condition,
\begin{equation}\label{e.10}
B^4+B^2C^2+2AB(AD+GC)-(GA-HC)^2+f(AD+GC)^2>0
\end{equation}

\section{The Penrose process}\label{III}
\begin{figure}[htbp]
\centering
\subfigure[]{\includegraphics[height=4cm,width=7cm]{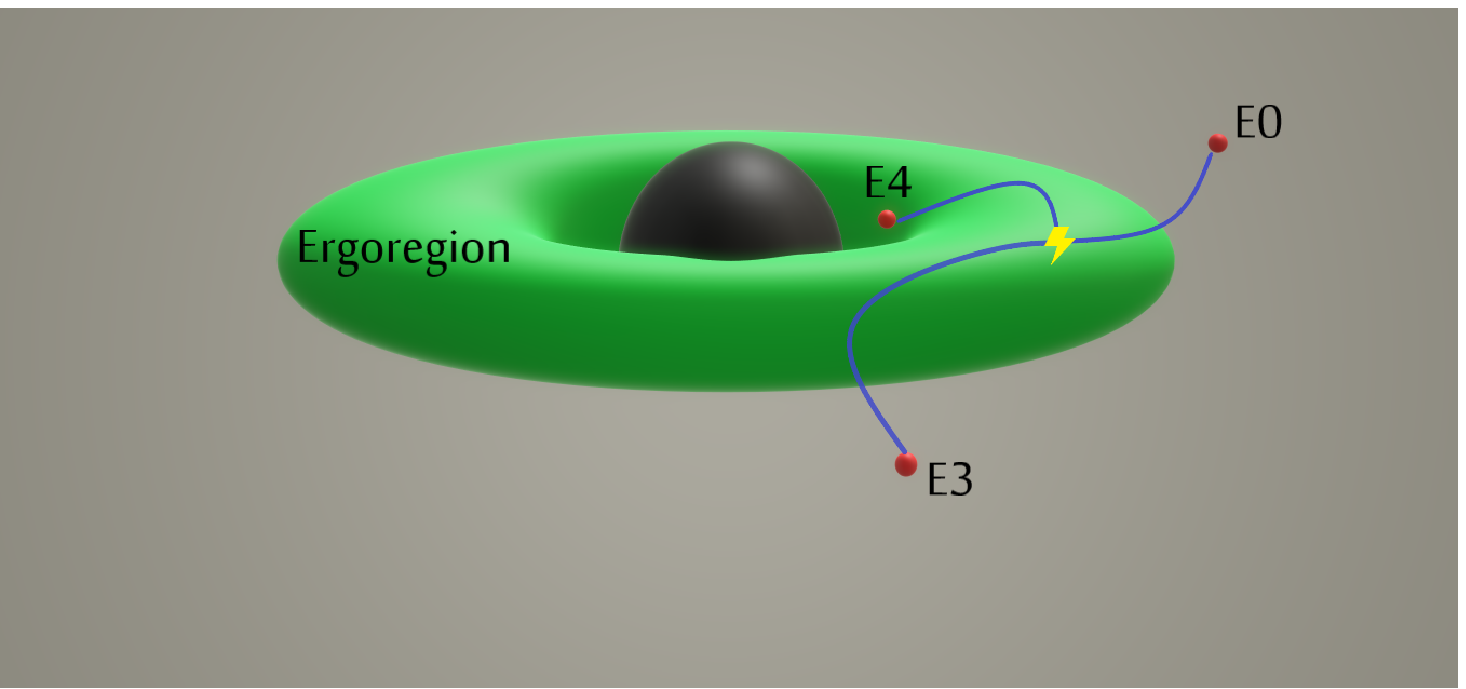}\label{fig01}}
\subfigure[]{\includegraphics[height=4cm,width=7cm]{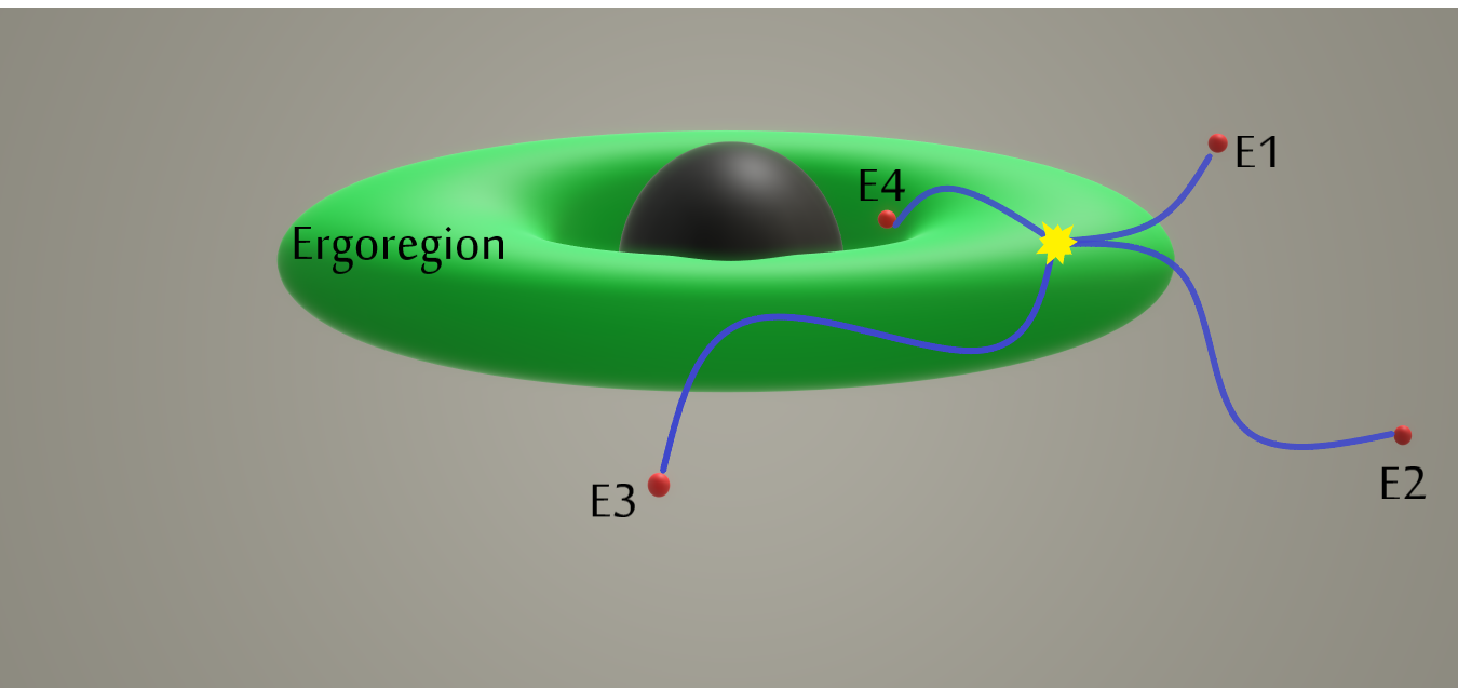}\label{fig02}}
\caption{In (a), the original Penrose process shows that the primary particle is divided into two other particles. In (b), the collision Penrose shows that the two primary particles are divided into two other particles.}
\label{fig1}
\end{figure} 
In the original Penrose process, a particle that enters the ergosphere region of a black hole splits into two particles. One particle falls into the black hole with negative energy, and the other escapes infinitely with positive energy. However, here, we consider the process of collision Penrose. According to it, two particles collide in the ergosphere, which results in two particles with positive and negative energy. Like the original state, a particle falls into a black hole with negative energy, and another escapes infinitely with positive energy. We assume 4-momentum and spin particles are conserved in this collision, i.e.,
\begin{equation}\label{e.11}
S_1+S_2=S_3+S_4,\hspace{30pt}P_1+P_2=P_3+P_4
\end{equation}
From which we can deduce the survival of energy and angular momentum,
\begin{equation}\label{e.12}
J_1+J_2=J_3+J_4,\hspace{30pt}E_1+E_2=E_3+E_4
\end{equation}
In this paper, only we study the collision of two massive particles. In this case, we consider the particle mass to be the same,
\begin{equation}\label{e.13}
m_1+m_2=m_3+m_4
\end{equation}
So we have,
\begin{equation}\label{e.14}
u^{(1)}_1+u^{(1)}_2=u^{(1)}_3+u^{(1)}_4,\hspace{30pt} s_1+s_2=s_3+s_4,
\end{equation}
The amount of angular momentum and energy of the particles determines their orbit around the black hole. There is a critical limit to these values, which we call $\Gamma_c=E_c/J_c$. Particles enter the black hole at $\Gamma<\Gamma_{c}$ orbits, and for $\Gamma>\Gamma_{c}$ orbits, it returns before the horizon. We consider the collision very close to the horizon $r_c\approx r_H$. According to (\ref{e.11}), the particle orbits can have different states. Including critical $\Gamma=E/J$, near-critical $\Gamma=(1+\xi)E/J$, and non-critical $\Gamma=(1+\sigma\epsilon+\gamma\epsilon^2)E/J$. We can extract the maximum energy from the black hole by applying these conditions to the equations (\ref{e.12}).

\section{The Kerr–Newman black hole}\label{IV}
The Kerr-Newman metric is the most common solution of the Einstein-Maxwell equations in space-time asymptotically flat, which explains space-time geometry near a charged rotating mass. This metric in Boyer Lindquist coordinate is,
\begin{equation}\label{e.15}
\begin{split}
ds^{2}=\frac{f(r)}{\rho^2}(dt-aSin^2\theta)^2+&\frac{\rho^2}{f(r)}dr^2+\rho^2d\theta^2+\frac{Sin^2\theta}{\rho^2}((r^2-a^2)^2d\phi-adt)^2,\\
f(r)&=r^2-2Mr+a^2+Q^2,\\
\rho^{2}&=r^2+a^2Cos^2\theta\\
r_\pm&=M\pm\sqrt{M-(a^2+Q^2)}
\end{split}
\end{equation}
$Q$, $a=J/M$, $M$, and $r_\pm$ are charge, angular momentum, mass, and the horizon radius of a black hole, respectively.
\subsection{The motion equations}
Its conserved quantity can be obtained by rewriting the Kerr-Newman metric as (\ref{e.1}) and using (\ref{e.7}).
\begin{equation}\label{e.16}
\begin{split}
\frac{E}{m}=&[\sqrt{\frac{r^2-2Mr+a^2+Q^2}{r^2-a^2}}+(\frac{(a^2+3r^2)(1+r)-2Q^2r}{2(r^4-a^4)})\sqrt{\frac{r^2-2Mr+a^2+Q^2}{r^4+a^2(2r+r^2-Q^2)}}a^2s]u_0\\
+&[\sqrt{\frac{r^4+a^2(2r+r^2-Q^2)}{r^2-a^2}}+\frac{2rs(Q^2+M(\frac{a^2}{r}-r)-a^4(\frac{1}{2}-\frac{1}{r})+2a^2+r^4)}{2(r^2-a^2)^2}]u_3,
\end{split}
\end{equation}
and,
\begin{equation}\label{e.17}
\begin{split}
\frac{J}{m}=&[-\frac{ars\sqrt{r^2-2Mr+a^2+Q^2}}{(r^2-a^2)^2}]u_0+[\frac{a^2}{\sqrt{(a^2-r^2)(r^4+a^2(2r+r^2-Q^2))}}+(\frac{a^2(M-2r)}{(a^2+r^2)^2}\\
+&\frac{r(Mr-Q^2)}{(a^2+r^2)^2}-\frac{2r^3a^4+a^6(1+r)}{2(a^2-r^2)(r^4+a^2(2r+r^2-Q^2))^\frac{3}{2}(a-\sqrt{r^4+a^2(2r+r^2-Q^2)})})s]u_3.
\end{split}
\end{equation}
Also, We can calculate the motion equations of spinning particles near the Kerr-Newman background by using (\ref{e.8}) and (\ref{e.15}).
\begin{equation}\label{e.18}
\begin{split}
P^{t}(r)=&\frac{m^2}{\rho\omega}\textbf{[}r^2aJ(2Mr-Q^2)-\frac{Q^2r^2sE}{m}(r^6+2a-\frac{J}{E})-r^3a^2E(2M\\
+&r-\frac{Q^2}{r})+\frac{sa^2}{m}(aE-J)(Q^2-Mr)+\frac{Mr^3s}{m}(J-3aE)\textbf{]},
\end{split}
\end{equation}
\begin{equation}\label{e.19}
\begin{split}
P^{r}=&\frac{\delta m^3r^2}{\omega}\textbf{[}2M-\frac{a^2+Q^2}{r}+\frac{(J-aE)^2(2Mr-Q^2)}{r^3m^2}-\frac{(J^2-a^2E^2)}{rm^2}+\frac{E^2r}{m^2}-r+(((\frac{Q^2}{r^3}-\frac{M}{r^2})\\
-&(J-aE)\frac{EQ^2}{r}+3EM)(J-aE)-\frac{EJr}{a})\frac{2sa}{r^2m^3}-\frac{\alpha s^4}{r^7m^4}(\frac{Q^2}{r}-M)^2+\frac{s^2a^2E^2}{r^4m^4}(\frac{Q^2}{r}-M)\\
+&(\frac{Q^2}{r^2}-\frac{M}{r}+2)\frac{s^2}{r^5m^2}(\frac{J^2}{m^2}(\frac{Q^2}{r}-M)^2-(Q^2-Mr)(1-\frac{M}{r}+\frac{Q^2}{r^2})\frac{2aEJ}{m^2}-2a^2(Q^2-Mr)\\
-&(Q^2-r^2-2Mr)(\frac{r^2E^2}{m^2}2(Q^2-Mr)))\textbf{]}^{\frac{1}{2}},
\end{split}
\end{equation}
Where $\delta=\pm$, $+$, and $-$ show the direction of a particle moving outward and inward, respectively, and
\begin{equation}\label{e.20}
\begin{split}
P^{\phi}(r)=&\frac{m^2}{\rho\omega}\textbf{[}aEr^2(Q^2-2Mr)+\frac{sEa^2}{m}(Q^2-Mr)+r^2Q^2m\\
+&\frac{asJ}{m}(Mr-Q^2)+r^3(r-2M)(\frac{sE}{m}-J)\textbf{]},
\end{split}
\end{equation}
$\rho$ and $\omega$ are,
\begin{equation}\label{e.21}
\begin{split}
\omega=&-Mm^{2}s^{2}r+M^{2}r^{4}+Q^{2}m^{2}s^{2},\\
\rho=&-2Mr+Q^{2}+a^{2}+r^{2}.
\end{split}
\end{equation}
\subsection{The constraints}
\begin{figure}[htbp]
	\centering
	\subfigure[]{\includegraphics[height=4cm,width=8cm]{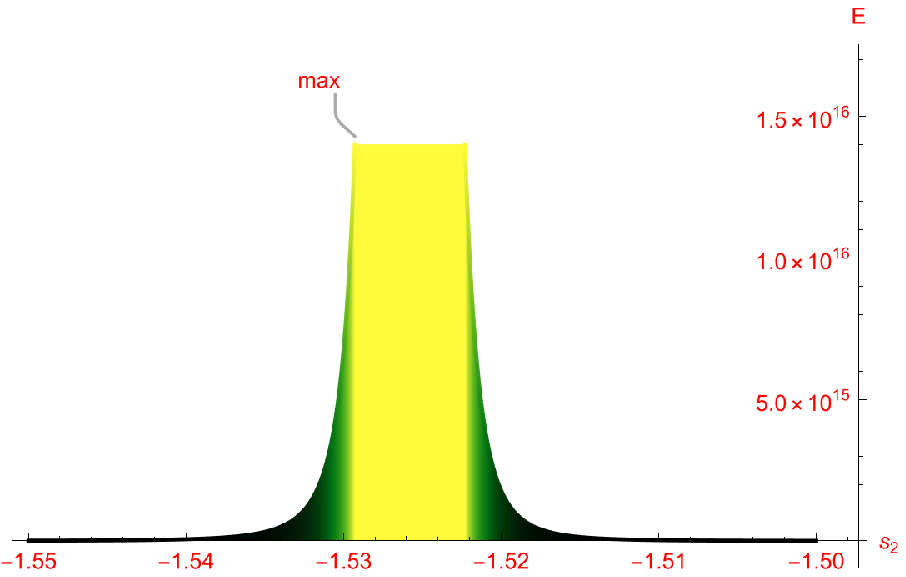}\label{fig03}}
	\subfigure[]{\includegraphics[height=4cm,width=8cm]{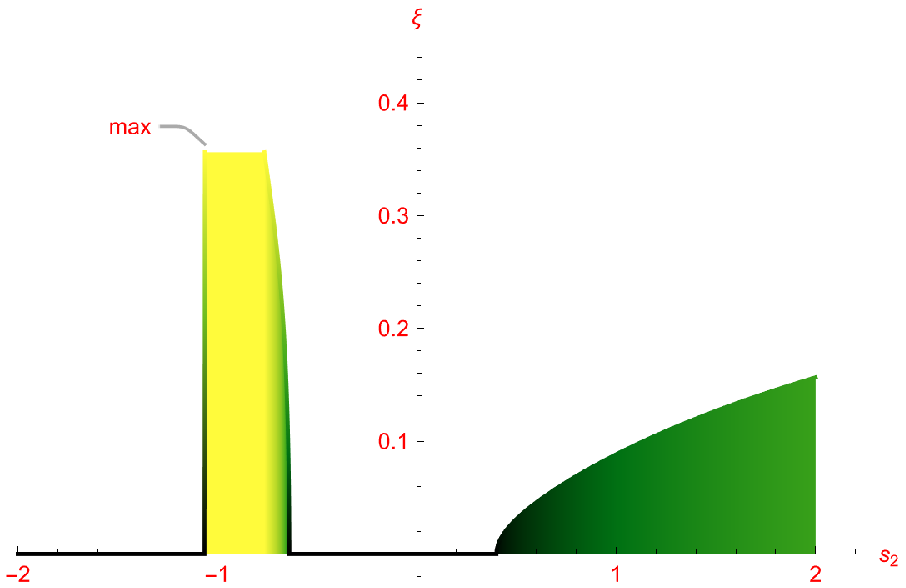}\label{fig04}}
	\caption{ In (a), permissible area for energy and spin according to which the particle can reach the horizon. In (b), the non-critical particles $0\leq\zeta\leq 0.4$ have a positive spin of $0.7\leq s\leq 1.7$.}
	\label{fig2}
\end{figure}

Before continuing, we examine three constraints.
\paragraph{\textbf{I.}} The first under consideration is the WGC condition. In charged rotating black holes, the extremal condition is $M^2=a^2+Q^2$. To avoid naked singularity, we must consider $M^2\geq a^2+Q^2$. As a result, we consider the best-case scenario, the near-extremal, and obtain the results.
\begin{equation}\label{e.22}
M^2=\lim_{\varepsilon \longrightarrow 0} (a^2+Q^2+\varepsilon)
\end{equation}
\paragraph{\textbf{II.}} The second constraint is on the particle orbit. We find the orbits of particles on the horizon by $P^{r}(r_H)\geq 0$. By introducing the compact parameter $\Gamma=J/E$ in the third section, we define a critical value for $\Gamma_{c}$, which indicates that no more particles can reach the horizon. The turning point $(\Gamma=\Gamma_{c})$ is in the horizon's radius, where the particle returns. Therefore, for the orbit of particles to reach the horizon, the necessary condition for energy and angular momentum is $\Gamma\leq\Gamma_{c}$. Can be obtained by (\ref{e.19}) and (\ref{e.22}) for $\varepsilon=0.1$ (the near-extremal),
\begin{equation}\label{e.23}
\Gamma_{c}=\frac{J_{c}}{E_{c}}=\frac{17.23+14.8s-2.04s^2-9.76(0.34-1.63s^2-1.24s^3+0.01s^4)^{\frac{1}{2}}}{6.14+5.55s+s^2},
\end{equation}
\paragraph{\textbf{III.}} The last constraint, 4-velocity, must be time-like, i.e., $v^av_a<0$, and must be off the event horizon $r\geq r_H$. We can get (\ref{e.10}) for Kerr-Newman black hole by equivalent placement (\ref{e.1}) and (\ref{e.15}), And then we get the third constraint using (\ref{e.23}).
\begin{equation}\label{e.24}
\begin{split}
E=&\frac{1}{b}[3.52+4.6s-\Gamma(4.18+0.56s)-(20.04+165.1s+336.13s^2-0.13s^3-0.24s^4+0.01s^5\\
&+0.02s^6-(31.65+194.45s-0.18s^4+264.64s^2-0.16s^3+0.01s^5+0.02s^6)\Gamma-(0.03s^4\\
&-12.45+0.03s^3-0.004s^5-52.06s^2-0.004s^6-50.94s)\Gamma^2)^{1/2}],\\
b=&11.08+67.37s+121.84s^2-\Gamma(20.86+80.82s+92.03s^2)+\Gamma^2(17.86s^2+18.99s+10.22)
\end{split}
\end{equation}
This energy constraint limits the spin for particles that fall from infinity. By examining $E\geq 1$, we obtain the solutions of (\ref{e.24}), $s=-1.56$, $-1.55$, $-0.92$, $-0.88$, $-0.06$, $0.026$, $0.51$, $0.69$. We can see the spins with the most energy by drawing (\ref{e.24}) Fig.\ref{fig2} (a).\\
The maximum spin is in $s_{max}=-1.525$. We examine the extraction energy in this spin. For $\Gamma\geq\Gamma_{c}$, a particle that falls from infinity returns to infinity at a turning point. So both directions of the particles are essential in the collision, and we will consider it In the following. If we calculate the time-like condition for non-critical particles $\Gamma=\Gamma_c(1+\xi)$, it gives us a limit on $\xi$. 
\begin{equation}\label{e.25}
\begin{split}
\xi=&\frac{1}{b}[473.33+7263.87s_2+43815.66s_2^2+128988.15s_2^3+182173.002s_2^4+97694.54s_2^5\\
&+14846.56s_2^6+0.12s_2^7+0.02s_2^8]-1,\\
b=&4887.28+64522.31s_2+323396.97s_2^2+762685.34s_2^3+814662.42s_2^4+268467.15s_2^5\\ 
&-45752.64s_2^6+0.27s_2^7-0.04s_2^8-(0.34-1.63s_2^2-1.24s_2^3+0.01s_2^4)^{\frac{1}{2}}(0.19s_2^6\\
&+2768.42+34170.93s_2+154165.54s_2^2+303648.72s_2^3+218897.37s_2^4+0.09s_2^5)
\end{split}
\end{equation} 

The permissible $\xi$ area is obtained by drawing (\ref{e.25}) Fig.\ref{fig2} (b).\\
\subsection{Particle collision}
This step investigates the collision of two massive particles in near-extremal conditions. By placing the condition (\ref{e.22}) in (\ref{e.15}) and calculating $f(r)= 0$, we get $M=r+\varepsilon$. Note that $\varepsilon$ tends to zero, so we think of it as a minor disorder and write it that way. Then using these conditions in (\ref{e.19}) for $r= 1$ and $\varepsilon=0.1$, we have,
\begin{equation}\label{e.26}
\begin{split}
P^{r}=&(\frac{\delta_{\pm}}{1.46-0.32s^2})\\
&\sqrt{32.3+33.43s+2.05s^2-0.009s^4-J(9.6+8.24s-1.13s^2)+J^2(0.62+0.56s+0.1s^2)}
\end{split}
\end{equation}
On the other hand, we select the particles according to section 3 so that the first critical particle is a non-critical second particle, and the third particle is near-critical, in which case most of the energy can be extracted from the black hole.
\begin{equation}\label{e.27}
\begin{split}
\frac{J_1}{m_1}=&\Gamma_c \frac{E_1}{m_1}\\
\frac{J_2}{m_2}=&\Gamma_c (1+\xi)\frac{E_2}{m_2}\\
\frac{J_3}{m_3}=&\Gamma_c (1+\sigma\epsilon+\gamma\epsilon^2+...)\frac{E_3}{m_3}\\
\frac{J_4}{m_4}=&\Gamma_c (\frac{E_1}{m_1}+\frac{E_2}{m_2}(1+\xi)-\frac{E_3}{m_3}(1+\sigma\epsilon+\gamma\epsilon^2+...)+...)
\end{split}
\end{equation}
We also consider $s_1=s_3$ and $s_2=s_4$ for the particles spin. Furthermore, as we mentioned in (\ref{e.13}), particles have equal masses, and the direction of the fourth particle that falls into the black hole is $\delta_4=\delta_2$. We calculate $P^{r}$ for each particle separately by applying all these conditions. By placing it in equation (\ref{e.11}), the energy of the third particle can be obtained in energy terms of the primary particles.
\begin{equation}\label{e.28}
E_3=\frac{-A\pm \sqrt{A^2-4BC}}{2B}
\end{equation}
Due to the large values of $A$, $B$, and $C$, we have included them in the appendix\ref{e.30}.
\subsection{Maximum energy efficiency}
\begin{figure}[htbp]
\centering
\subfigure[]{\includegraphics[height=6cm,width=8cm]{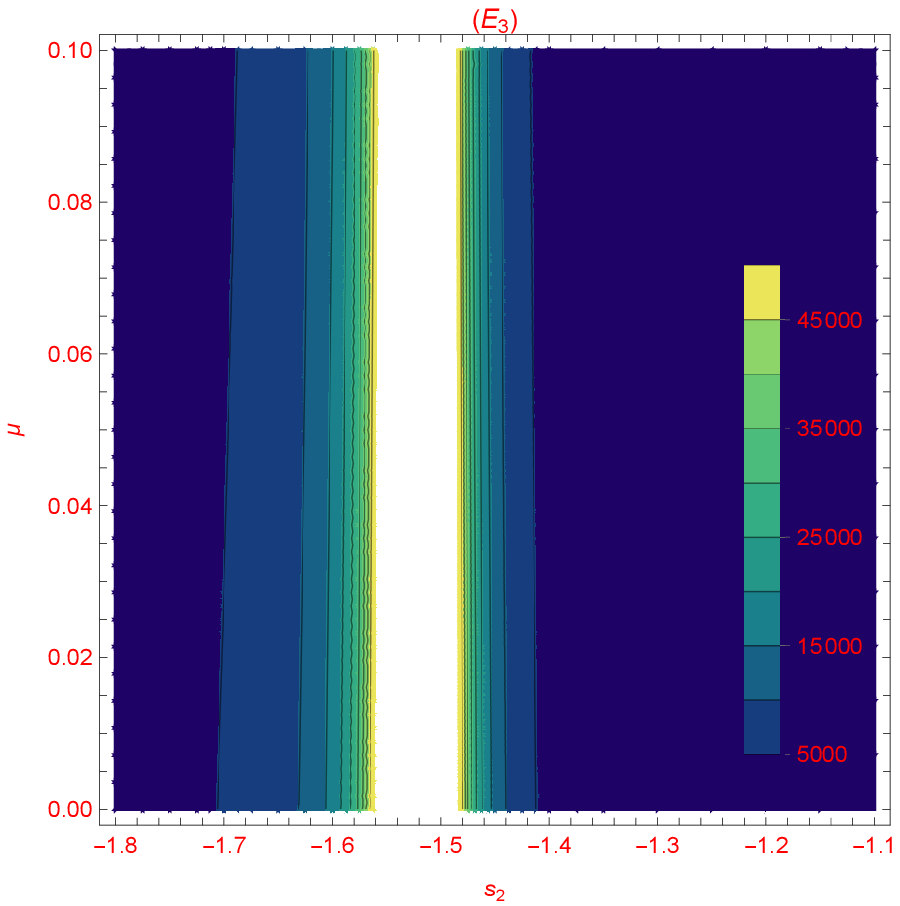}\label{fig05}}
\subfigure[]{\includegraphics[height=6cm,width=8cm]{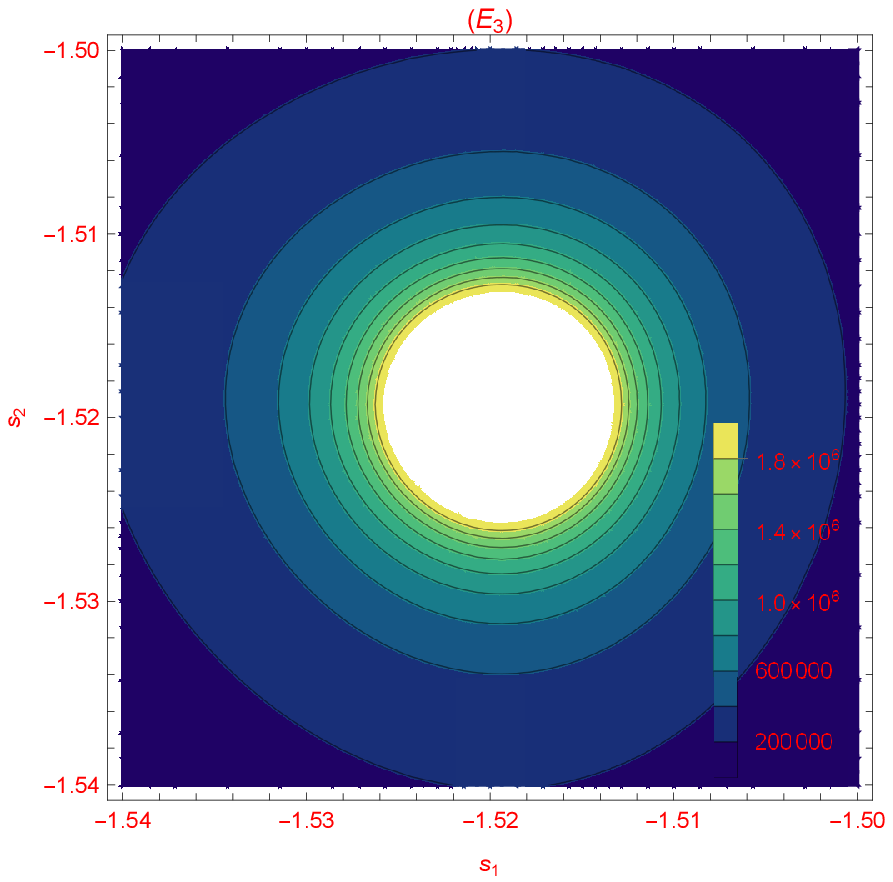}\label{fig06}}
\caption{ In (a), the third particle energy contour map for $s_1$ and $s_2$. In (b), The contour map of the maximal efficiency $E_3$ for $s_1$ and $s_2$.}
\label{fig3}
\end{figure}

In the last step, we calculate the maximum energy efficiency of the third particle in terms of the primary particles. We have $E_1/m_1\geq1$ and $E_2/m_2\geq1$ because particles come from infinity. As we said, we considered the second particle to be non-critical. According to Fig.\ref{fig2} (b), we have $0\leq\zeta\leq 0.4$ for $E_2/m_2=1$. As a result, we fix the second particle on $E_2/m_2=1$. We consider the energy of the first particle as arbitrary, $E_1=1$. This value well indicates the rate of increase of particle energy scattered near the Kerr-Newman black hole. We also consider the direction of the particles as $\delta_1=+1$, $\delta_2=\delta_4=-1$, $\delta_3=-1$ (This means that the third particle has the most energy when it exits the black hole and escapes to infinity). So we have two constraints, $\delta_3=-1$ and $s_3=s_1$, for the orbit of the third particle. It is also $E_3\geq E_{3,c}$ because the particles return to infinity in state $\Gamma\geq\Gamma_{c}$. Now we examine $E_3$ in terms of $\mu=(\sigma\epsilon+\gamma\epsilon^2+...)$. We fix spin $s_1(=s_1)$ on $s_{max}=-1.525$ as shown in Fig.\ref{fig1} and draw $E_3$ according to $\mu$ and $s_2$, the contour map of Fig.\ref{fig3}.\\
According to Fig.\ref{fig3}, the closer the energy is to interval $-1.48\leq s_2\leq-1.56$. Energy increases to $E_3=45000m_3$ and are $0\leq\mu\leq0.1$ variables at $s=-1.48,-1.56$ (This energy increase is the same for all $\mu$). As a result, we fix $\mu=0$ and plot the third particle energy $E_3$ in terms of $s_1(=s_3)$ and $s_2(=s_4)$, the contour map of Fig.\ref{fig3} (a).\\
The maximum amount of energy is in the range beyond $-1.526\leq s_1\leq-1.519$ and $-1.526\leq s_2\leq-1.513$, equal to $E_3=1800000m_3$. Also, the maximal efficiency equals,
\begin{equation}\label{e.29}
\eta=\dfrac{E_3/m_3}{E_1/m_1+E_2/m_2}=9\times10^5.
\end{equation}
This is the amount of energy for particles with a spin of $-1.54\leq s\leq-1.5$.
\subsection{The check of other particles}
In the last step, we examine the particles with different spins in our Penrose collision.\\
\begin{figure}[htbp]
	\centering
	\subfigure[]{\includegraphics[height=3.2cm,width=5.5cm]{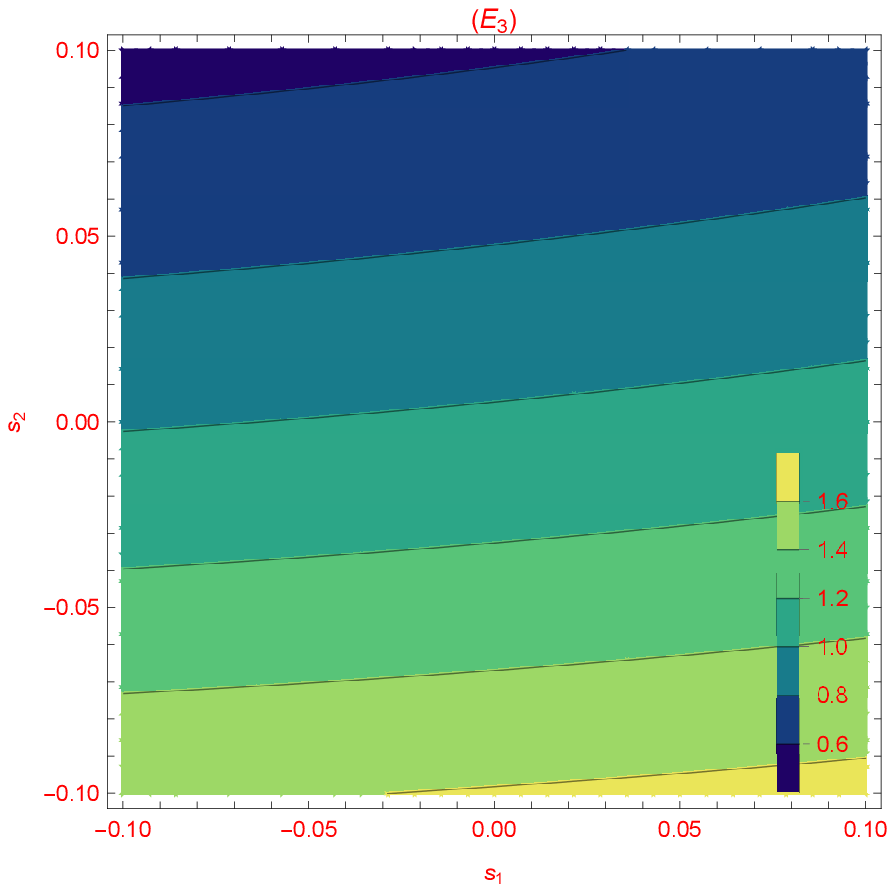}\label{fig07}}
	\subfigure[]{\includegraphics[height=3.2cm,width=5.5cm]{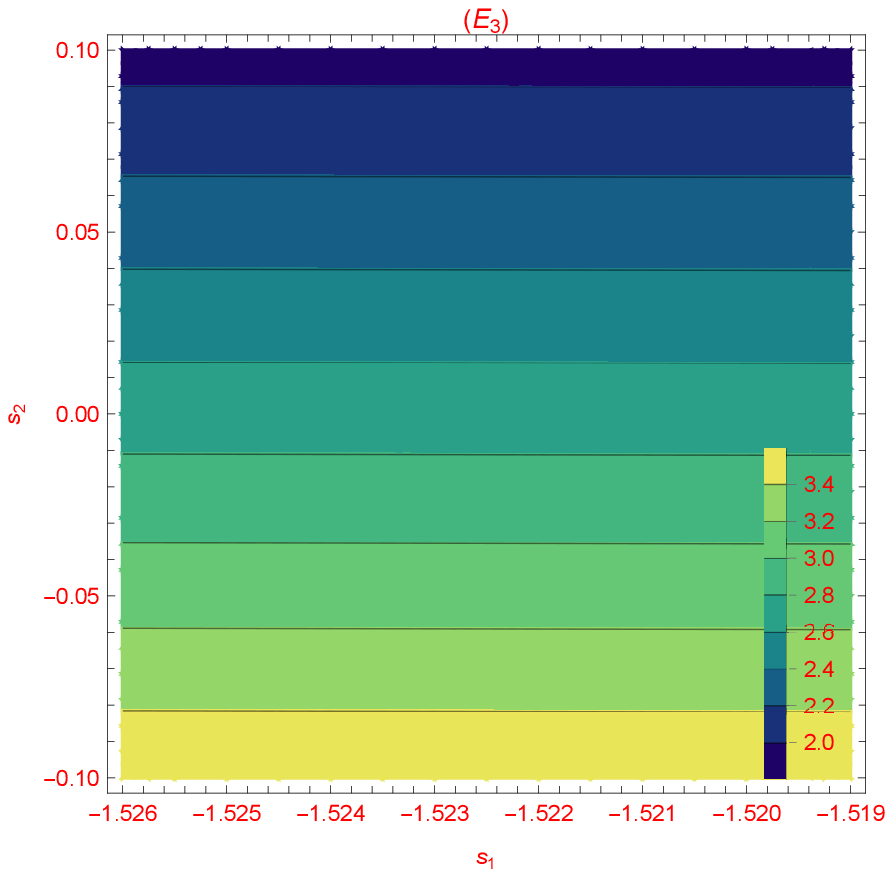}\label{fig08}}
	\subfigure[]{\includegraphics[height=3.2cm,width=5.5cm]{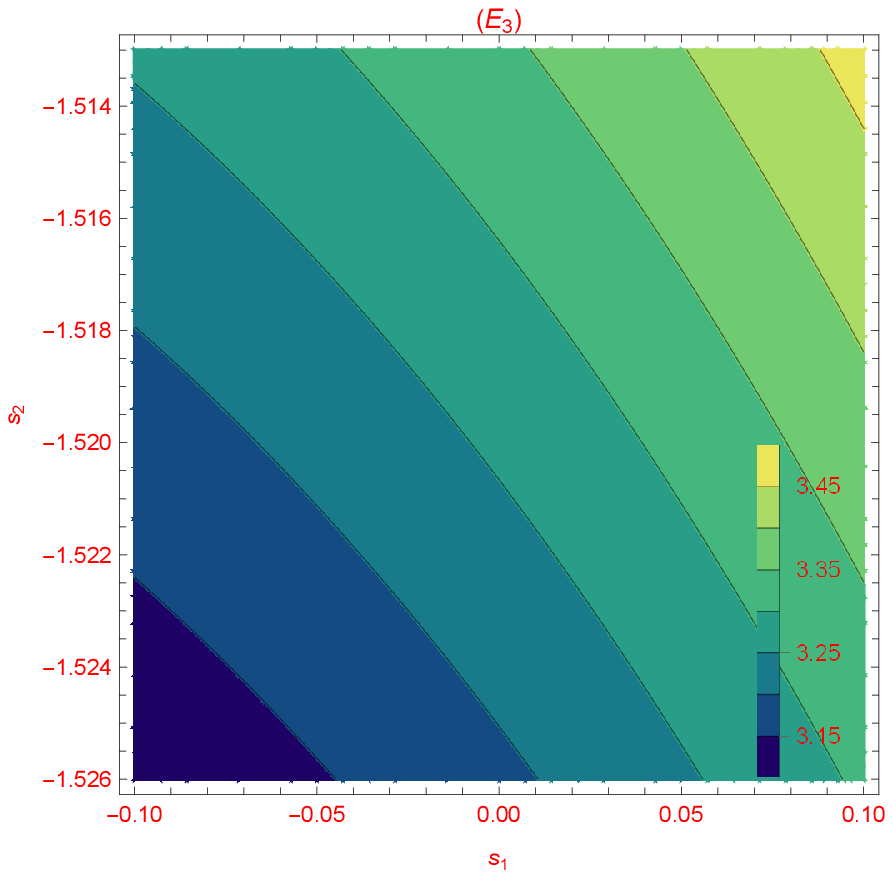}\label{fig09}}
	\subfigure[]{\includegraphics[height=3.2cm,width=5.5cm]{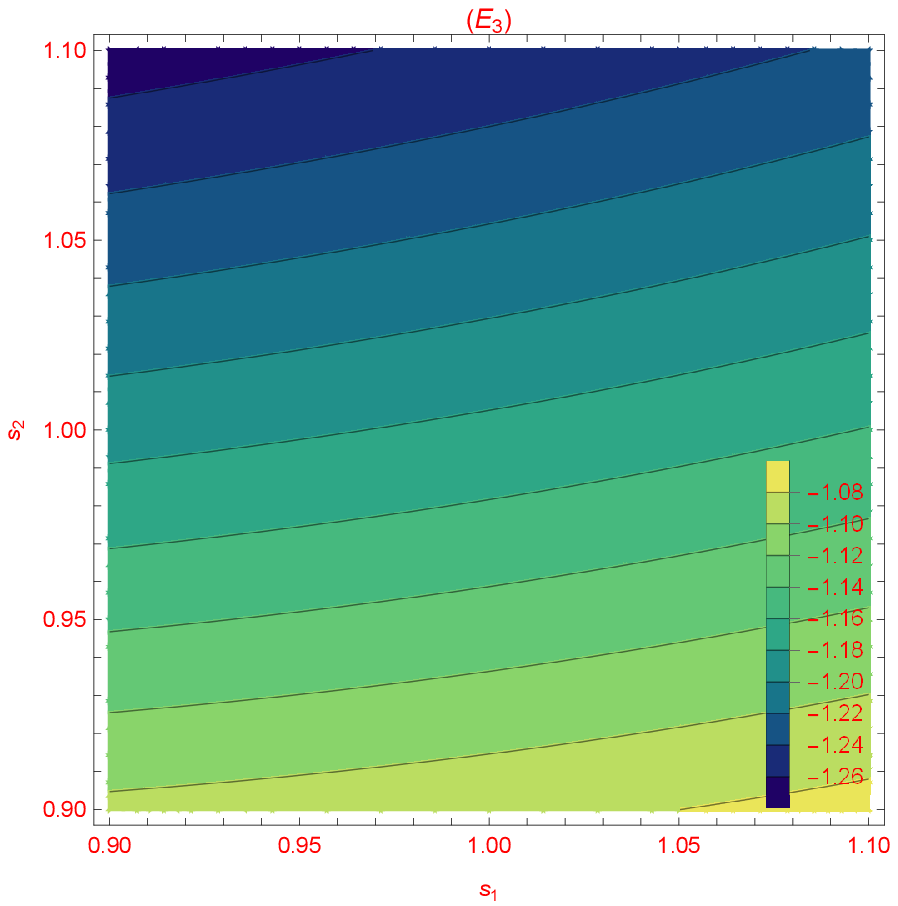}\label{fig10}}
	\subfigure[]{\includegraphics[height=3.2cm,width=5.5cm]{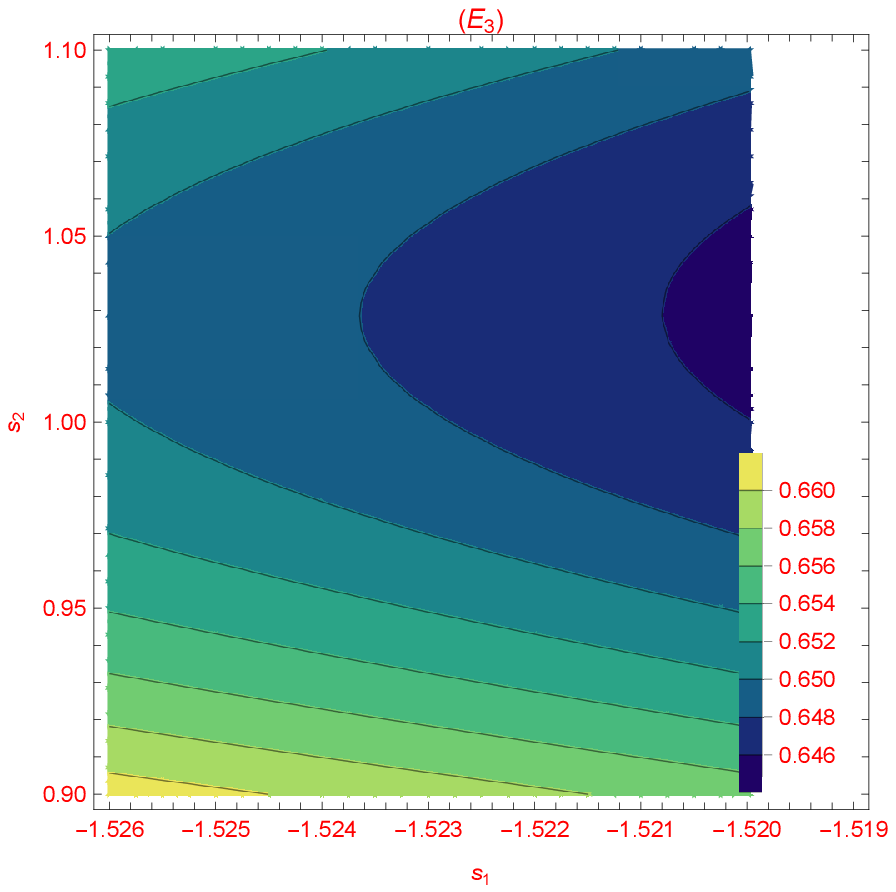}\label{fig11}}
	\subfigure[]{\includegraphics[height=3.2cm,width=5.5cm]{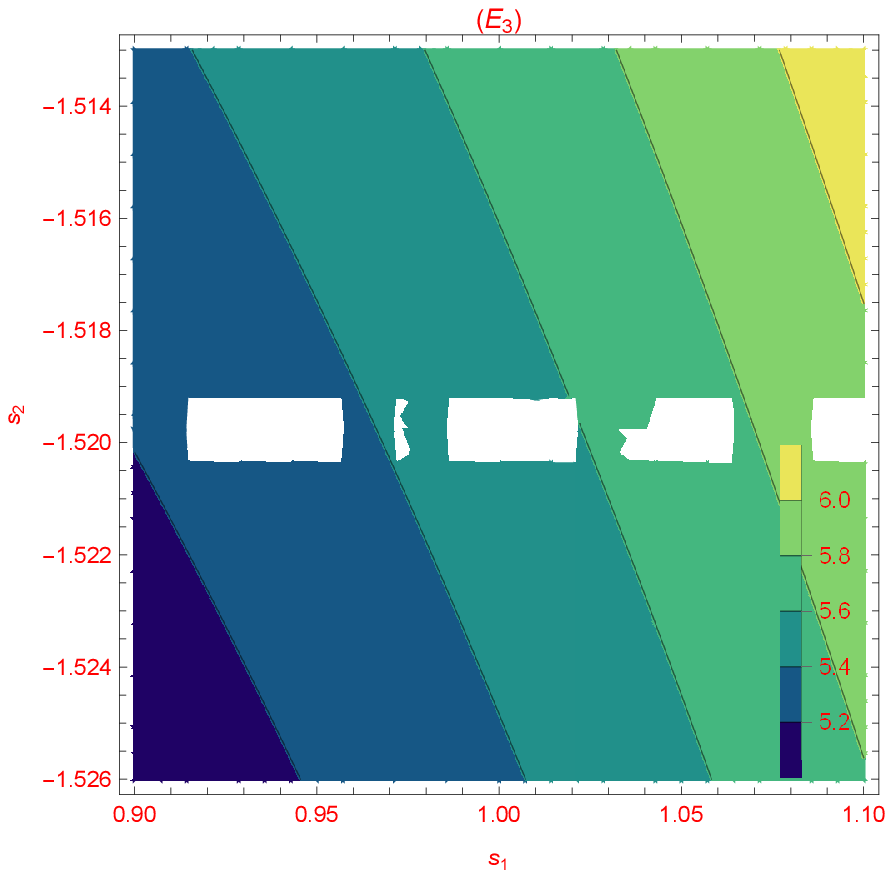}\label{fig12}}
	\subfigure[]{\includegraphics[height=3.2cm,width=5.5cm]{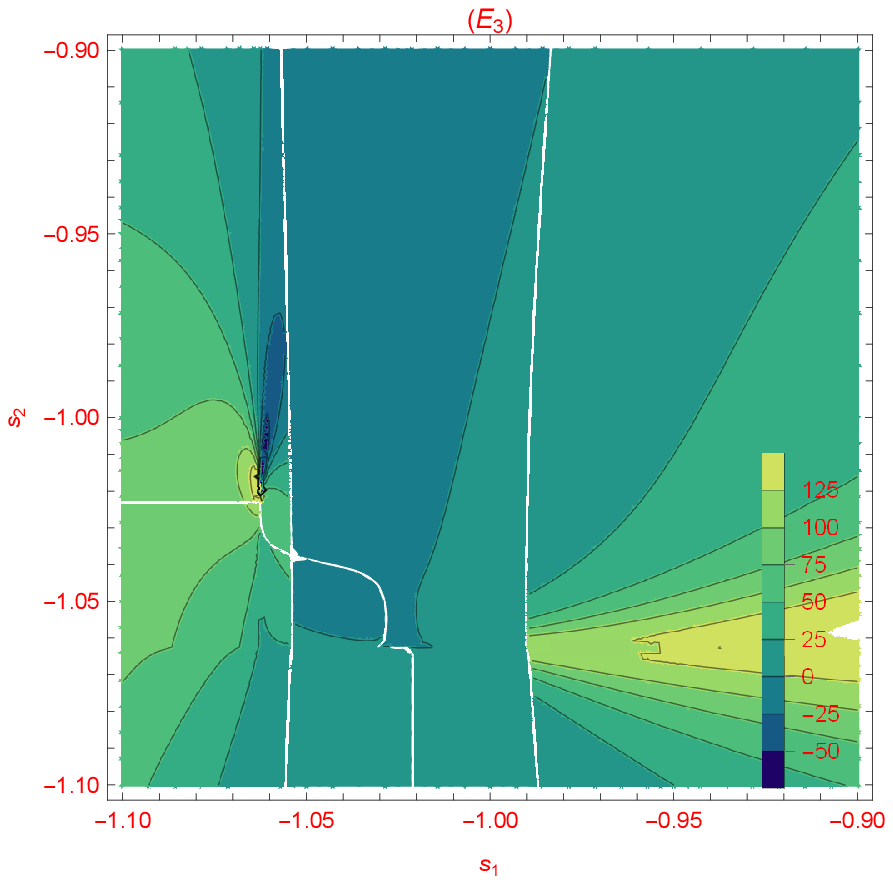}\label{fig13}}
	\subfigure[]{\includegraphics[height=3.2cm,width=5.5cm]{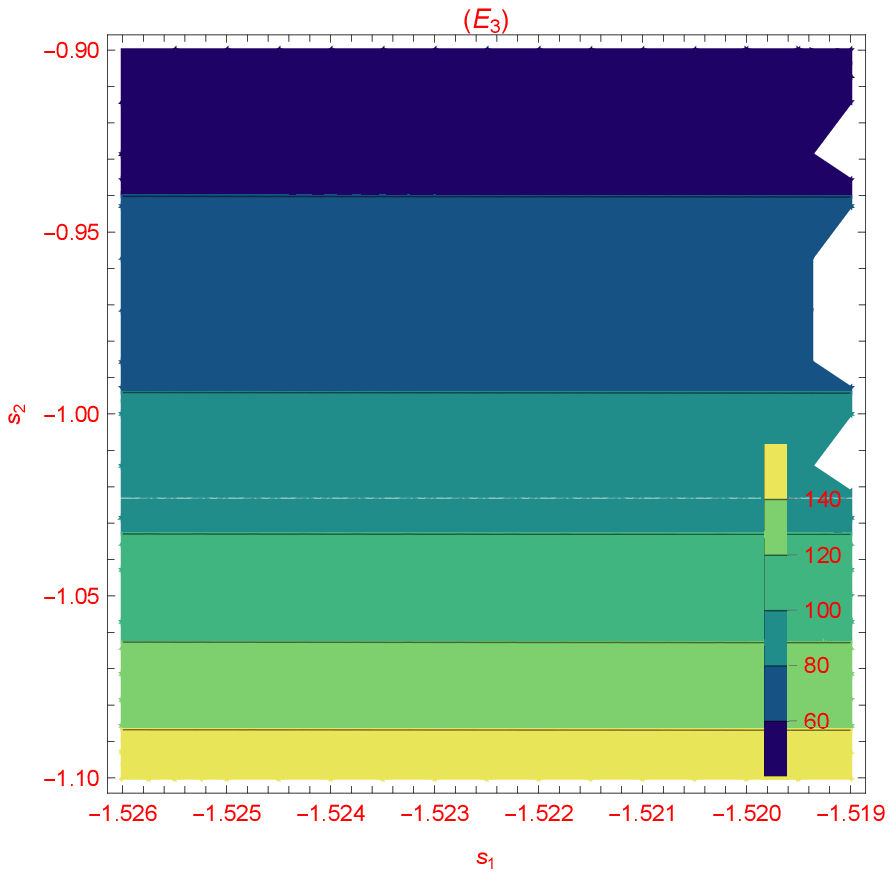}\label{fig14}}
	\subfigure[]{\includegraphics[height=3.2cm,width=5.5cm]{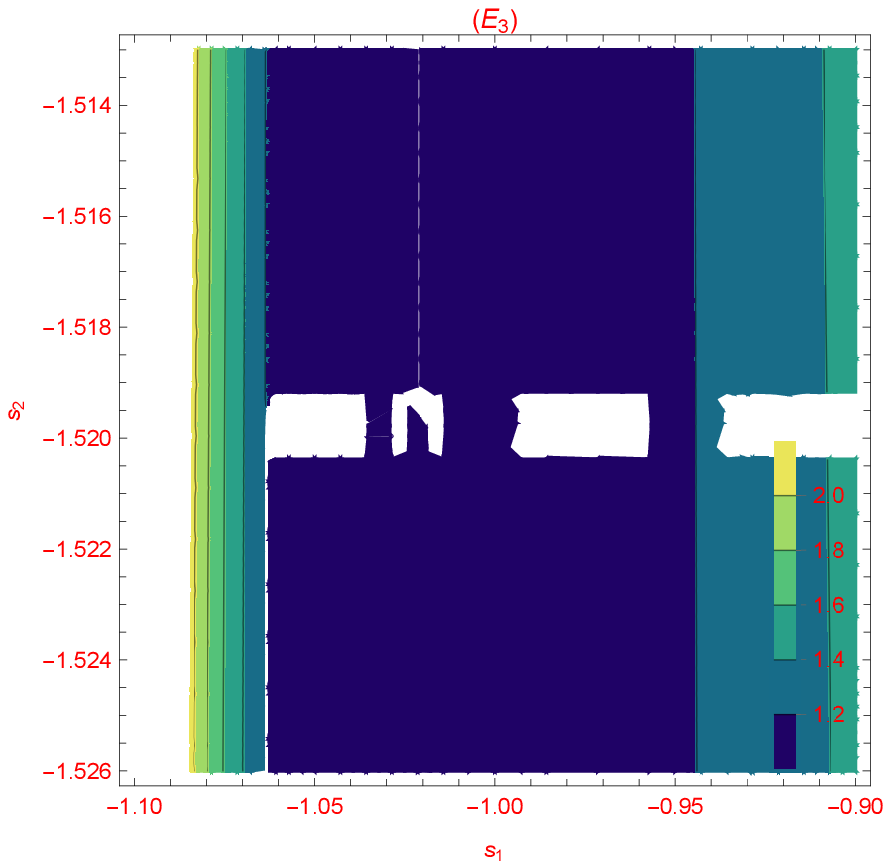}\label{fig15}}
	\subfigure[]{\includegraphics[height=3.2cm,width=5.5cm]{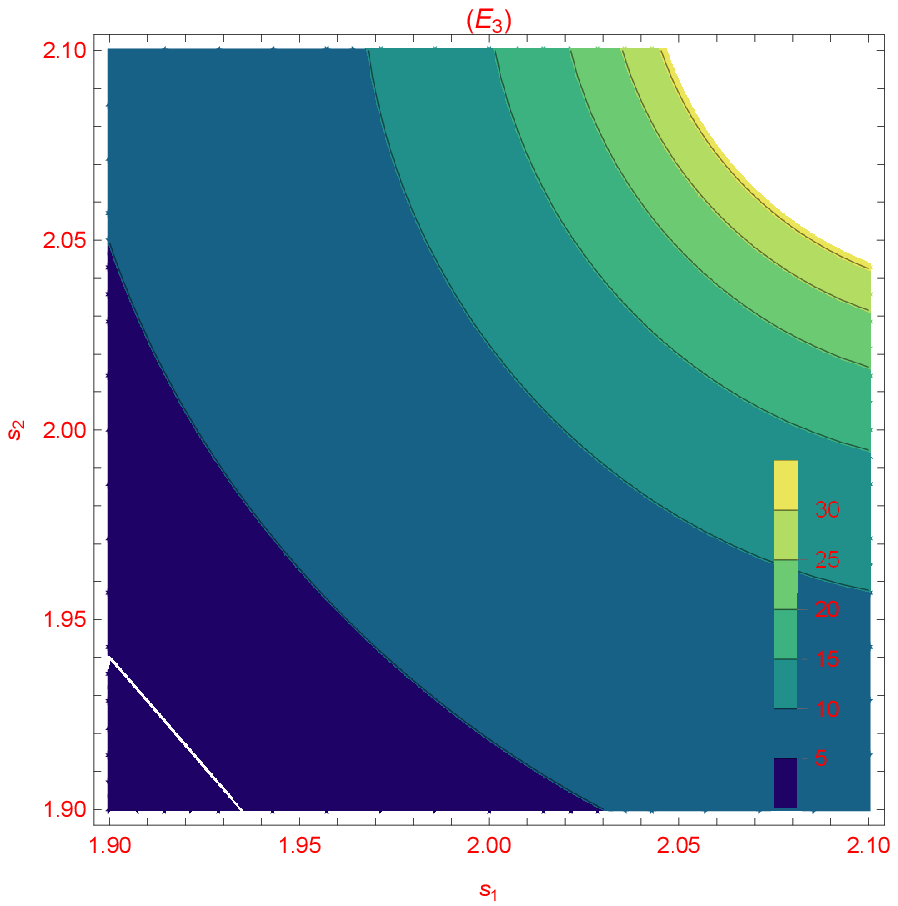}\label{fig16}}
	\subfigure[]{\includegraphics[height=3.2cm,width=5.5cm]{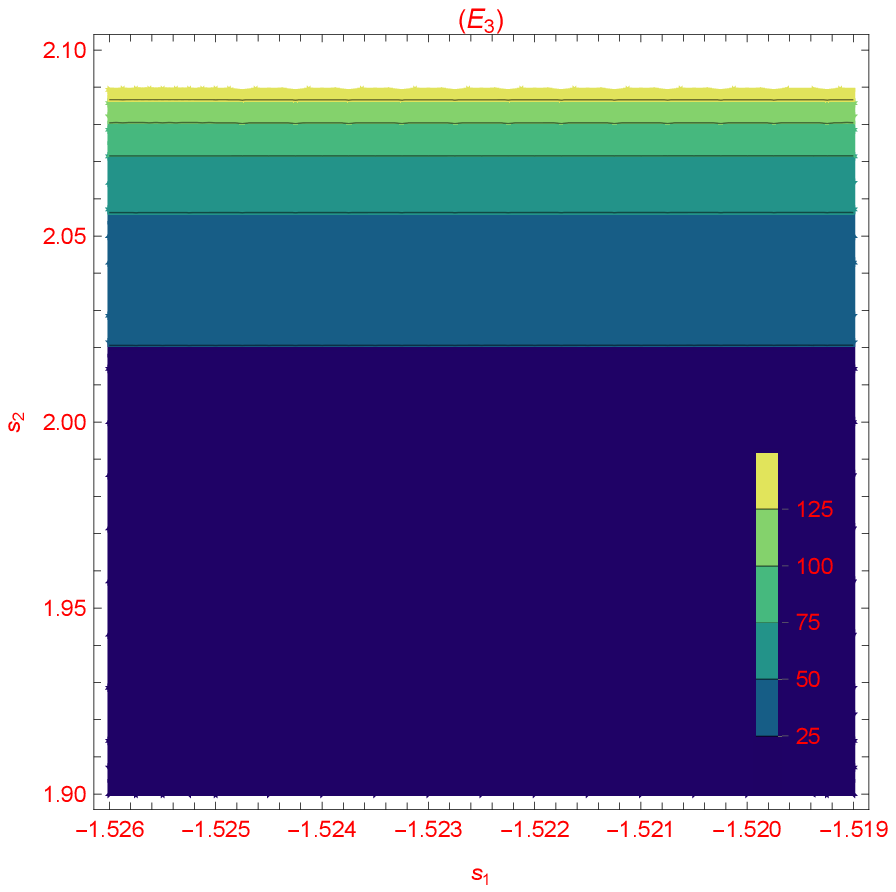}\label{fig17}}
	\subfigure[]{\includegraphics[height=3.2cm,width=5.5cm]{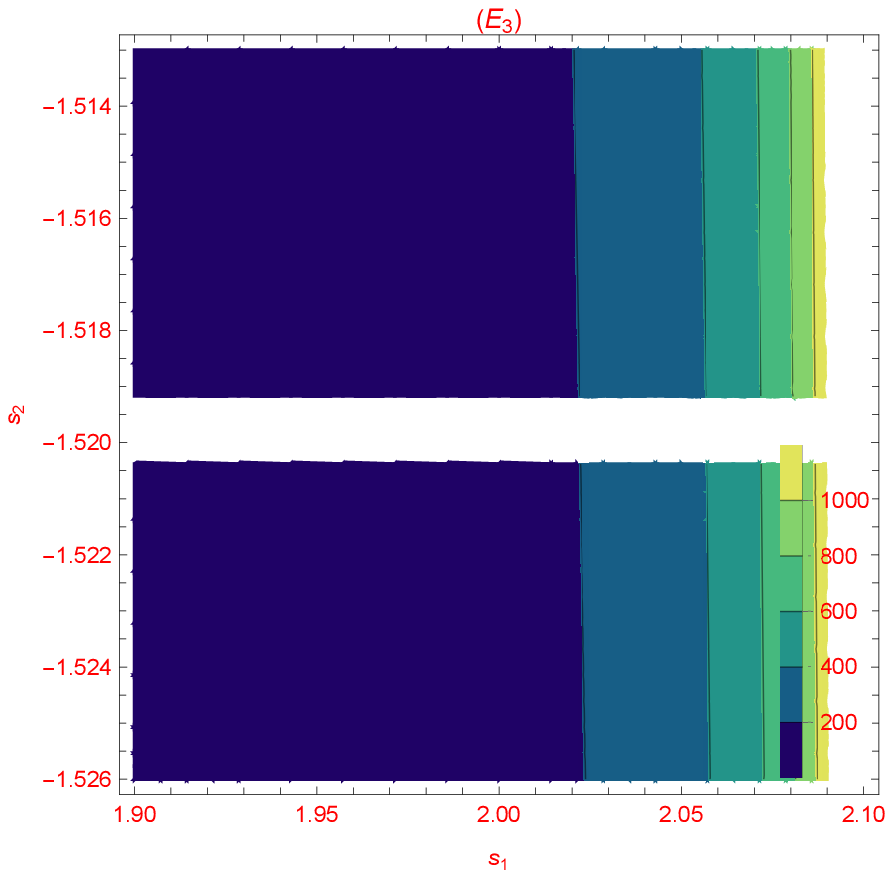}\label{fig18}}
	\subfigure[]{\includegraphics[height=3.2cm,width=5.5cm]{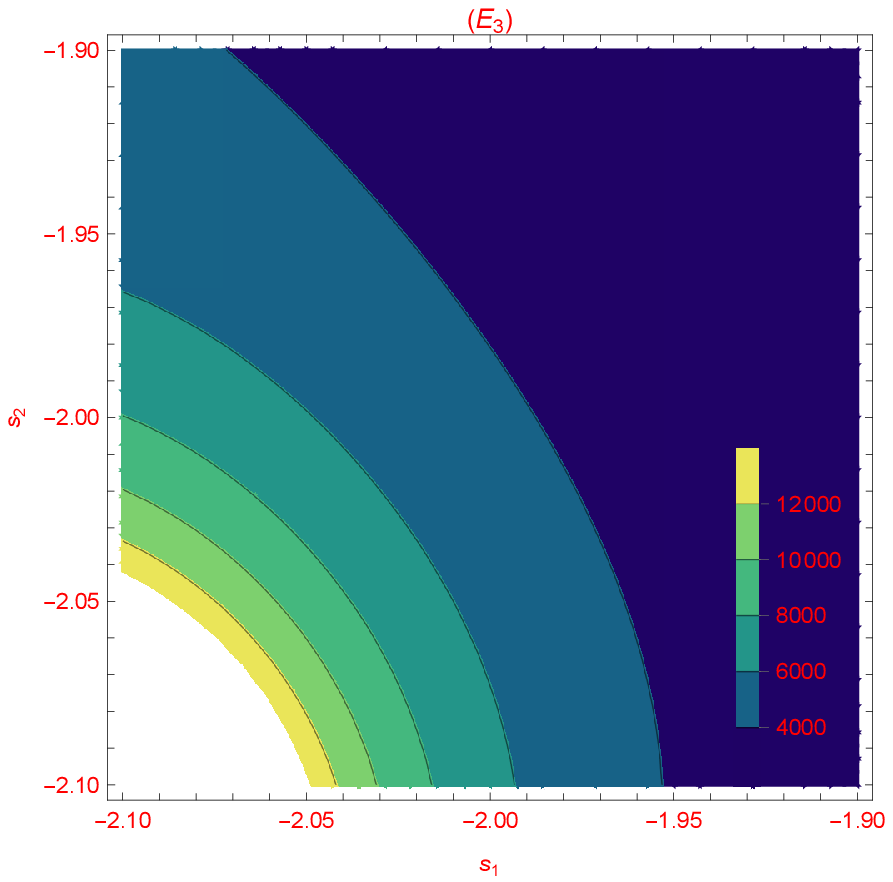}\label{fig19}}
	\subfigure[]{\includegraphics[height=3.2cm,width=5.5cm]{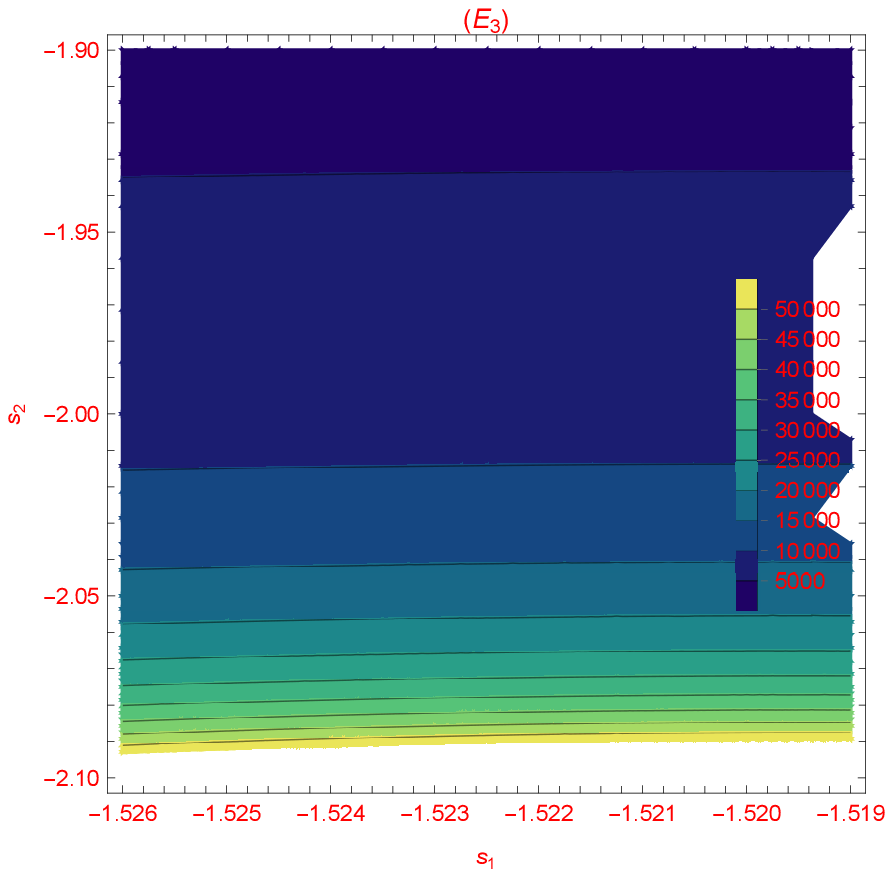}\label{fig20}}
	\subfigure[]{\includegraphics[height=3.2cm,width=5.5cm]{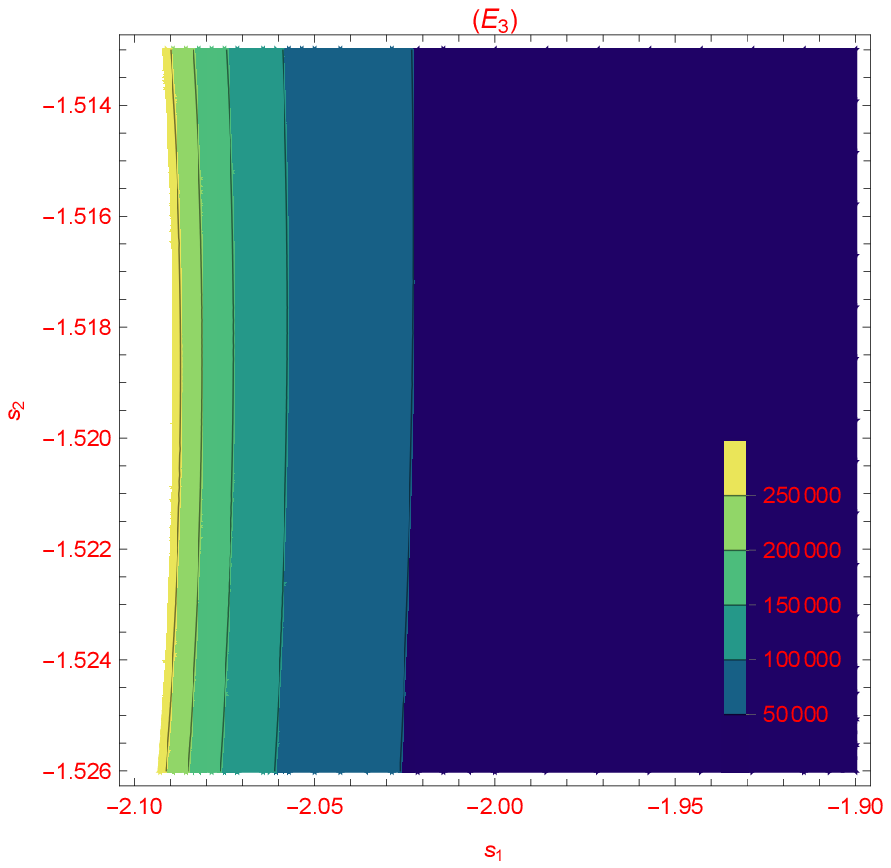}\label{fig21}}
	\caption{The check of bosons in the Penrose collision.}
	\label{fig4}
\end{figure} 

\begin{table}[tbp]
\centering
\begin{tabular}{|c|c|c|c|c|}
\hline
       &$check$  &$s_{3=1}\textbf{=}s_{4=2}$   &${(s_{3=1})}_{max}$          &${(s_{4=2})}_{max}$          \\
\hline
\hline
$1$    &$s=0$    &$-0.1\leq s_{3=4}\leq0.1$    & $-1.526\leq s_3\leq-1.519$  &$-0.1\leq s_3\leq0.1$        \\
	   &         &                             &$-0.1\leq s_4\leq0.1$        &$-1.526\leq s_4\leq-1.513$   \\
\hline
       &$E_3$    &$1.6 m_3$    $(a)$           &$3.4 m_3$  $(b)$             &$3.45 m_3$   $(c)$           \\
\hline
\hline
$2$    &$s=+1$   &$0.9\leq s_{3=4}\leq1.1$     &$-1.526\leq s_3\leq-1.519$   &$0.9\leq s_3\leq1.1$         \\
       &         &                             &$0.9\leq s_4\leq1.1$         &$-1.526\leq s_4\leq-1.513$   \\
\hline
       &$E_3$    &$-1.08 m_3$  $(d)$           &$0.66 m_3$    $(e)$          &$6 m_3$      $(f)$           \\
\hline
\hline
$3$    &$s=-1$   &$-1.1\leq s_{3=4}\leq-0.9$   &$-1.526\leq s_3\leq-1.519$  &$-1.1\leq s_3\leq-0.9$       \\
       &         &                             &$-1.1\leq s_4\leq -0.9$     &$-1.526\leq s_4\leq-1.513$   \\
\hline
       &$E_3$    &$125 m_3$  $(g)$             &$140 m_3$   $(h)$           &$2 m_3$  $(i)$               \\
\hline
\hline
$4$    &$s=+2$   &$1.9\leq s_{3=4}\leq2.1$     & $-1.526\leq s_3\leq-1.519$ &$1.9\leq s_3\leq2.1$         \\
       &         &                             &$1.9\leq s_4\leq2.1$        &$-1.526\leq s_4\leq-1.513$   \\
\hline
       &$E_3$    &$30 m_3$  $(j)$              &$125 m_3$   $(k)$           &$1000 m_3$   $(l)$           \\
\hline
\hline
$5$    &$s=-2$   &$-2.1\leq s_{3=4}\leq-1.9$   &$-1.526\leq s_3\leq-1.519$ &$-2.1\leq s_3\leq-1.9$        \\
       &         &                             &$-2.1\leq s_4\leq-1.9$     &$-1.526\leq s_4\leq-1.513$    \\
\hline
       &$E_3$    &$12\times 10^{3} m_3$  $(m)$ &$50\times 10^{3}m_3$$(n)$  &$250\times 10^{3} m_3$ $(o)$  \\
\hline
\end{tabular}
\caption{\label{tab1}Check of bosons in the Penrose collision.}
\end{table}
First, we check the boson particles. According to Fig.\ref{fig4} and table\ref{tab1}:\\
\textbf{1.} In the zero spin mode, it becomes $E_3=1.6m_3$ for both particles with $s=0$ (a). For a case where the output particle has a maximum spin value (b), $E_3=3.4m_3$, and if a particle falling into a black hole has a maximum spin (c),$E_3=3.45m_3$. In fact, in the two cases (b) and (c), there is no significant difference in the amount of extractive energy. Moreover, it becomes difficult and impossible to distinguish between the two.\\
\textbf{2.} In $s=+1$, we have an unacceptable value of $E_3=-1.08m_3$ (d). Nevertheless, for cases where one of the particles has a maximum spin (e and f), the values of extraction energy are acceptable, these values are equal to $E_3=0.66m_3$ and $E_3=6m_3$.\\
\textbf{3.} In $s=-1$, the extractive energy increases to a significant amount of $E_3=125m_3$ (g), and when $s_{max}$ in the collision occurs (h), it increases to $E_3=140m_3$, but when $s_{max}$ falls into the black hole (i), it decreases to $E_3=2m_3$.\\
\textbf{4.} We also study the graviton. In the case, the amount of extraction energy for particles with $s=2$ is $E_3=30m_3$ (g). For the case where $s_{max}$ comes out of the black hole, it becomes $E_3=125m_3$ (k), and when $s_{max}$ falls inside the black hole, and $s=+2$ comes out, it is $E_3=1000m_3$ (l).\\
\textbf{5.} In the last case, we consider the boson $s=-2$. The highest extraction energy is in this spin. For equal spin mode (m), we have $E_3=12\times 10^{3}m_3$. We have a prominent upward trend for the last two cases that brings the energy to $E_3=50\times 10^{3}m_3$ and $E_3=250\times 10^{3} m_3$. These values can help us observe gravitons and their effects in the future.
\begin{figure}[htbp]
	\centering
	\subfigure[]{\includegraphics[height=4cm,width=5.5cm]{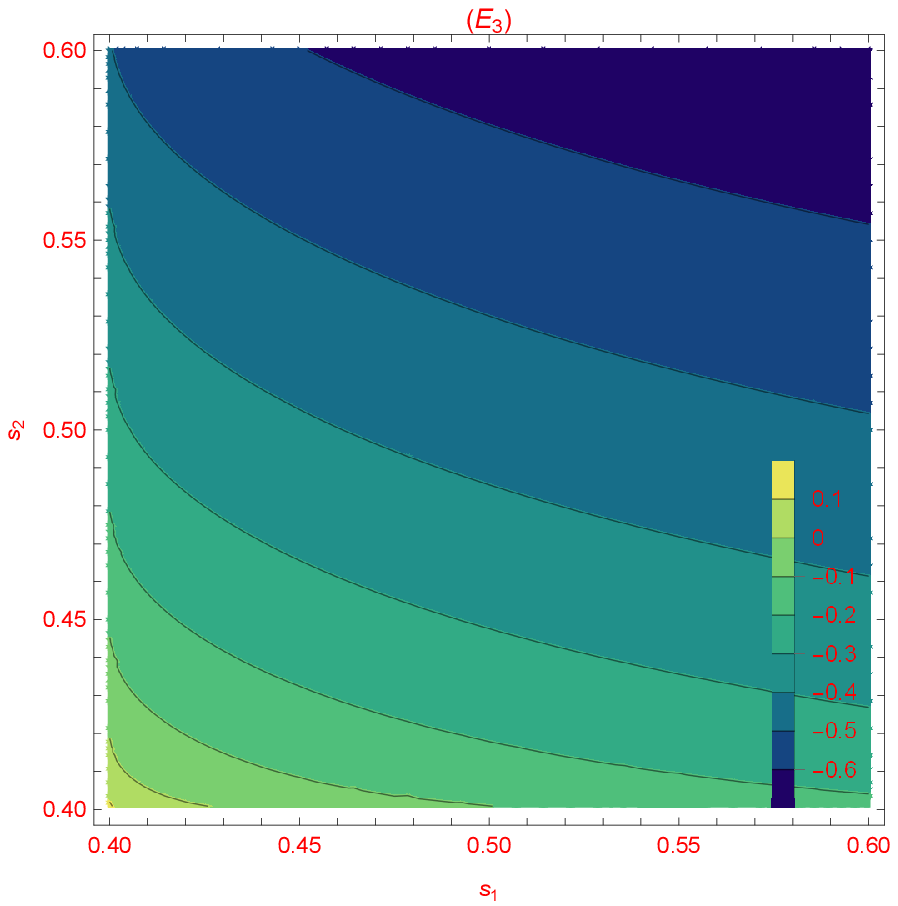}\label{fig22}}
	\subfigure[]{\includegraphics[height=4cm,width=5.5cm]{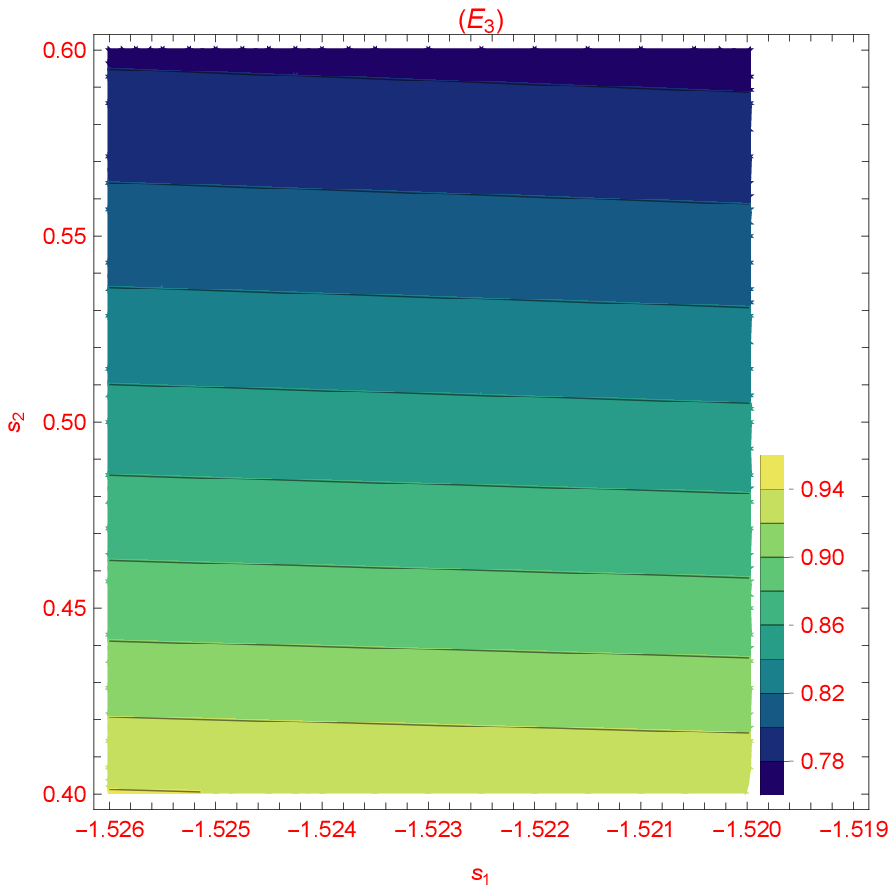}\label{fig23}}
	\subfigure[]{\includegraphics[height=4cm,width=5.5cm]{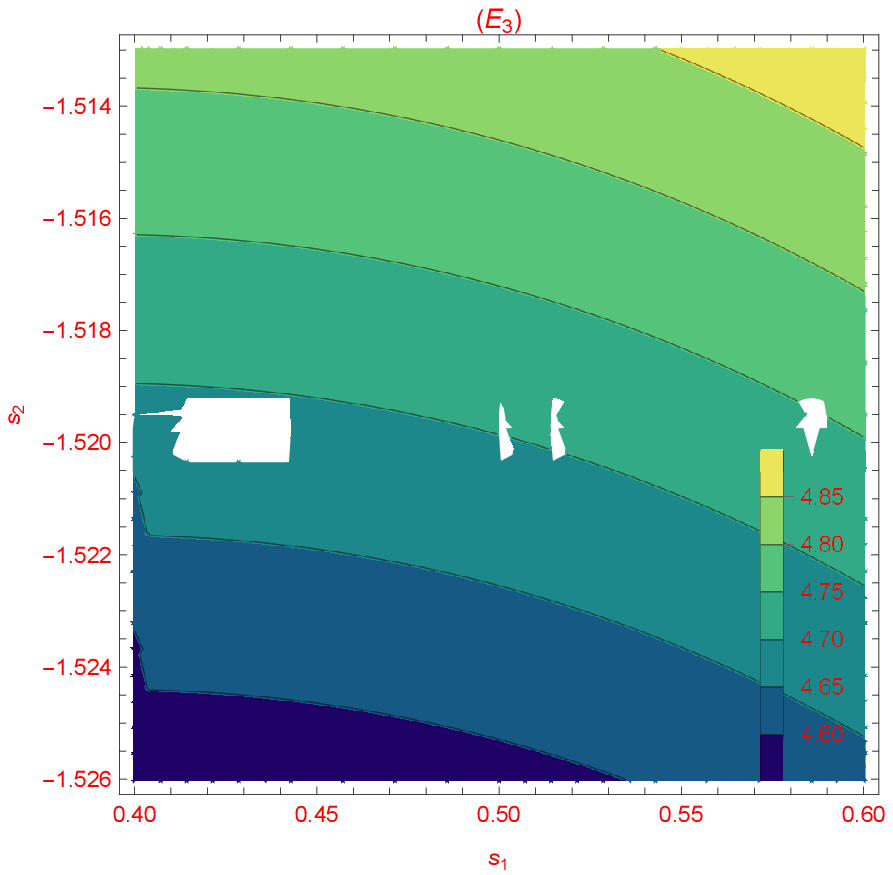}\label{fig24}}
	\subfigure[]{\includegraphics[height=4cm,width=5.5cm]{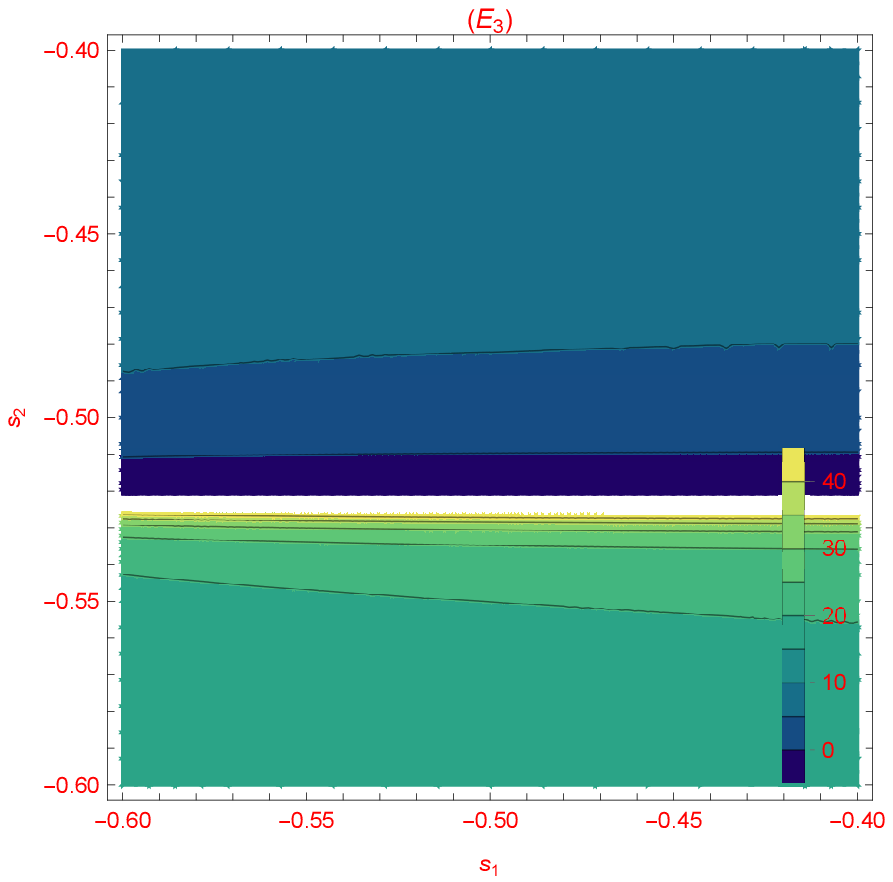}\label{fig25}}
	\subfigure[]{\includegraphics[height=4cm,width=5.5cm]{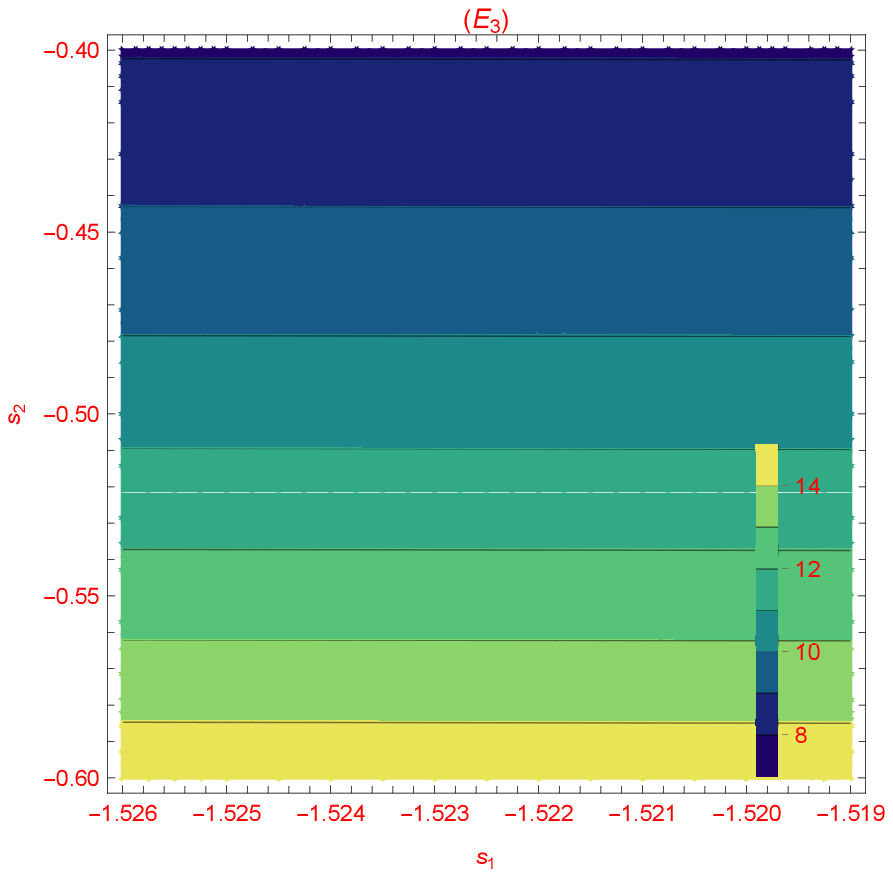}\label{fig26}}
	\subfigure[]{\includegraphics[height=4cm,width=5.5cm]{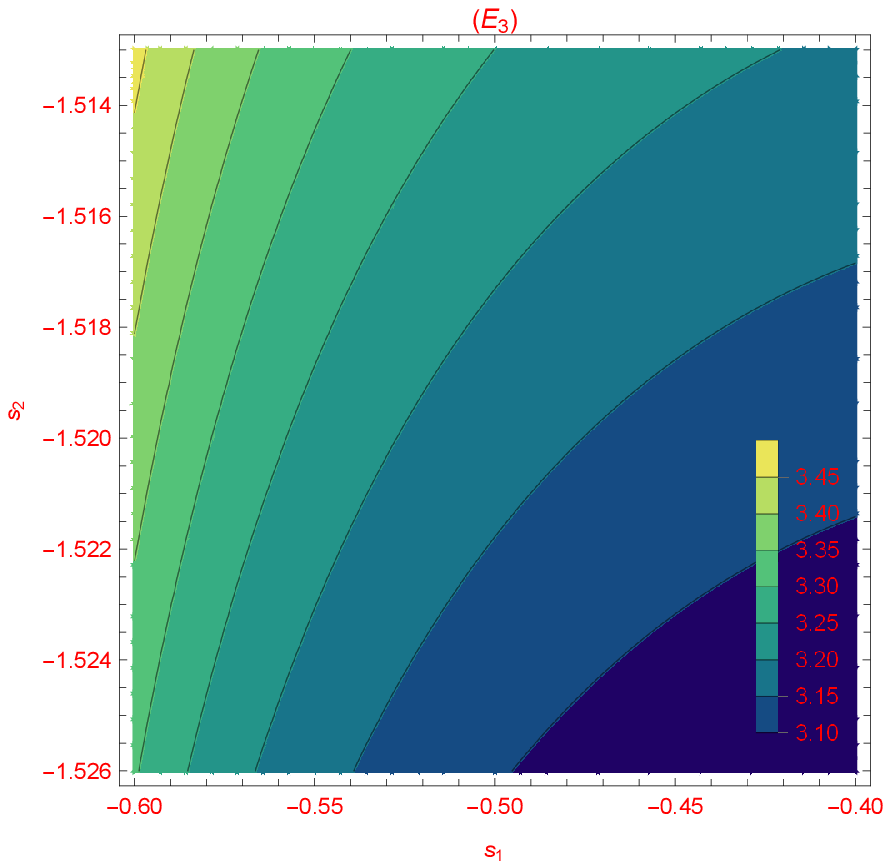}\label{fig27}}
	\subfigure[]{\includegraphics[height=4cm,width=5.5cm]{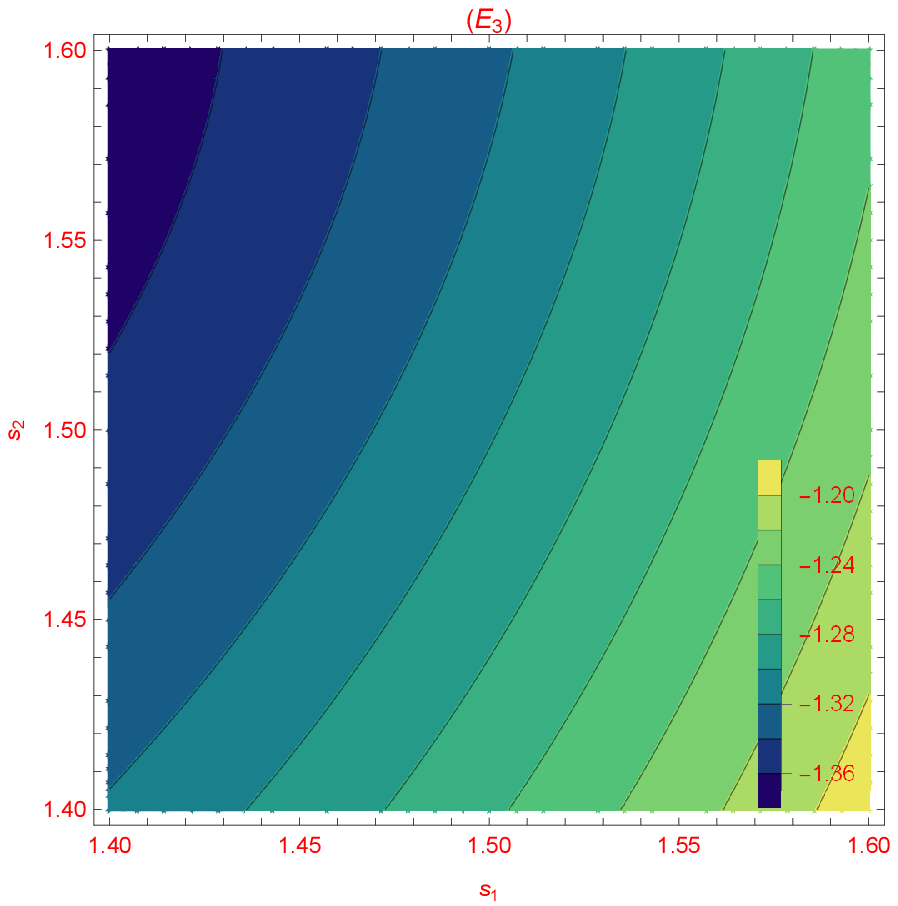}\label{fig28}}
	\subfigure[]{\includegraphics[height=4cm,width=5.5cm]{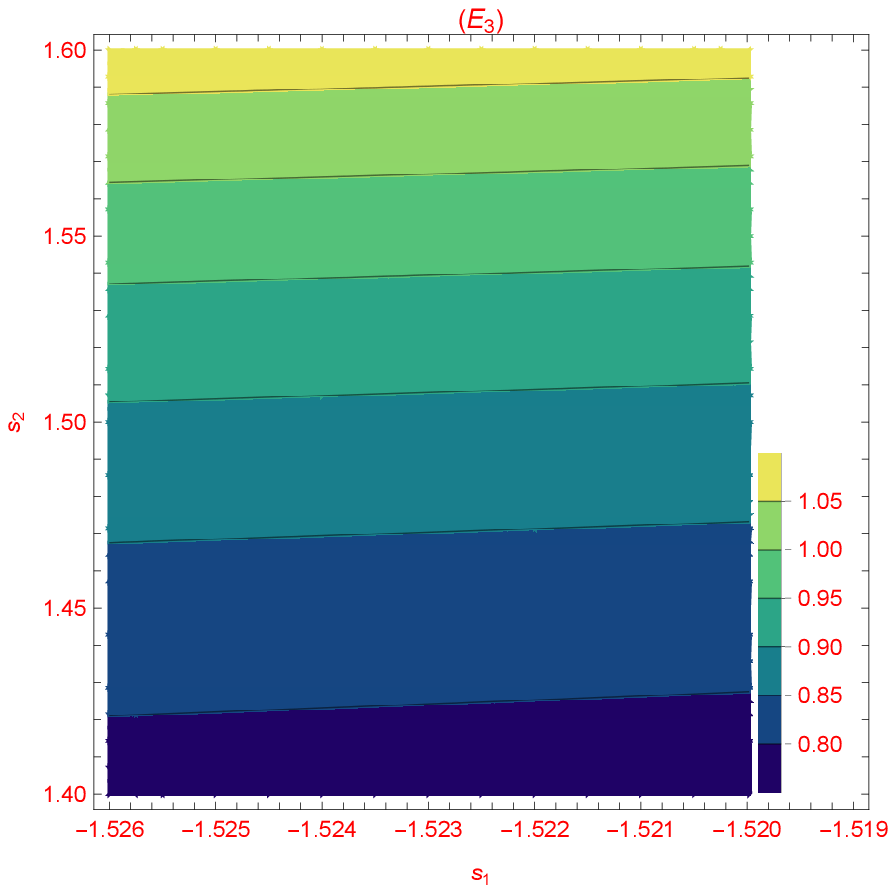}\label{fig29}}
	\subfigure[]{\includegraphics[height=4cm,width=5.5cm]{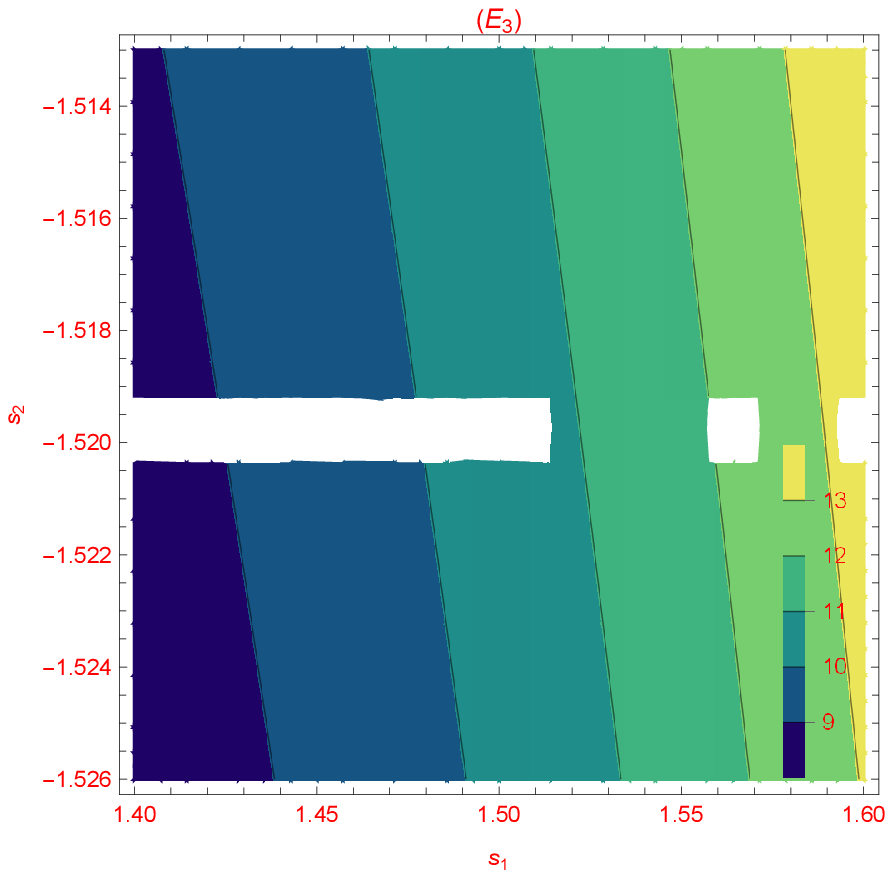}\label{fig30}}
	\subfigure[]{\includegraphics[height=4cm,width=5.5cm]{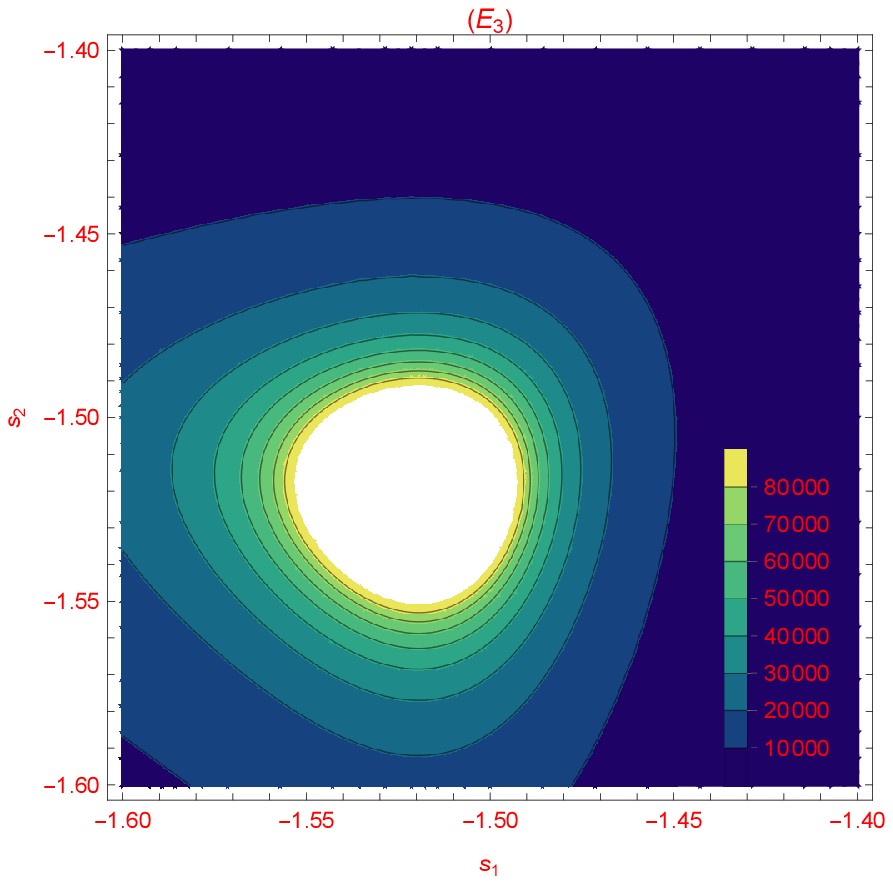}\label{fig31}}
	\subfigure[]{\includegraphics[height=4cm,width=5.5cm]{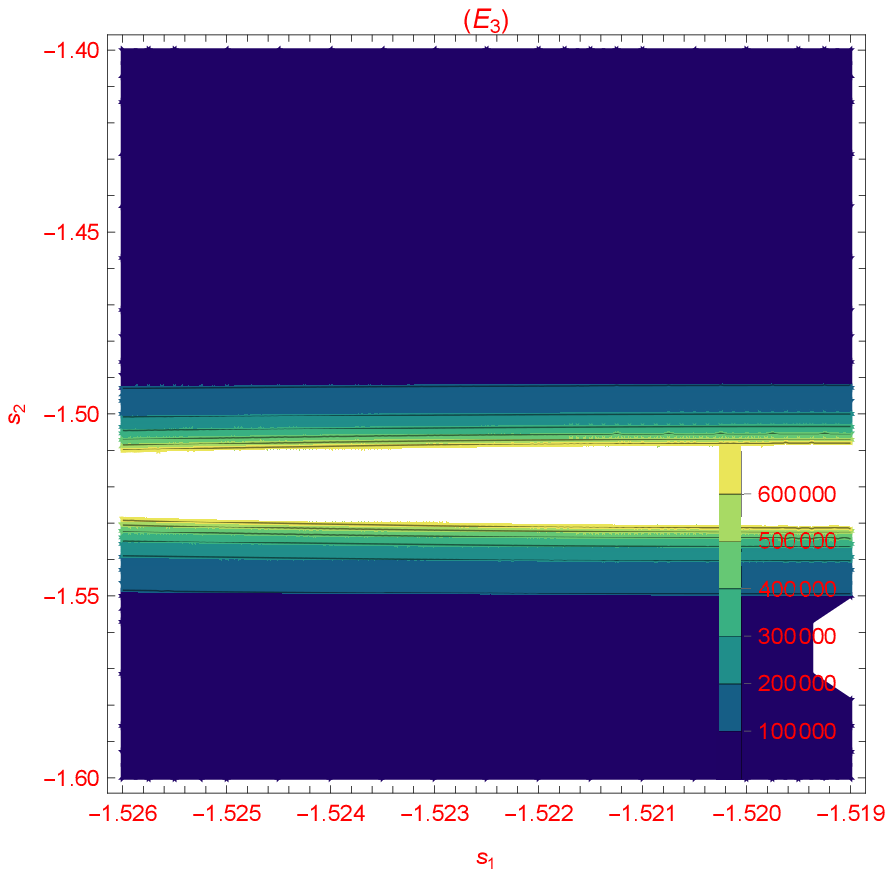}\label{fig32}}
	\subfigure[]{\includegraphics[height=4cm,width=5.5cm]{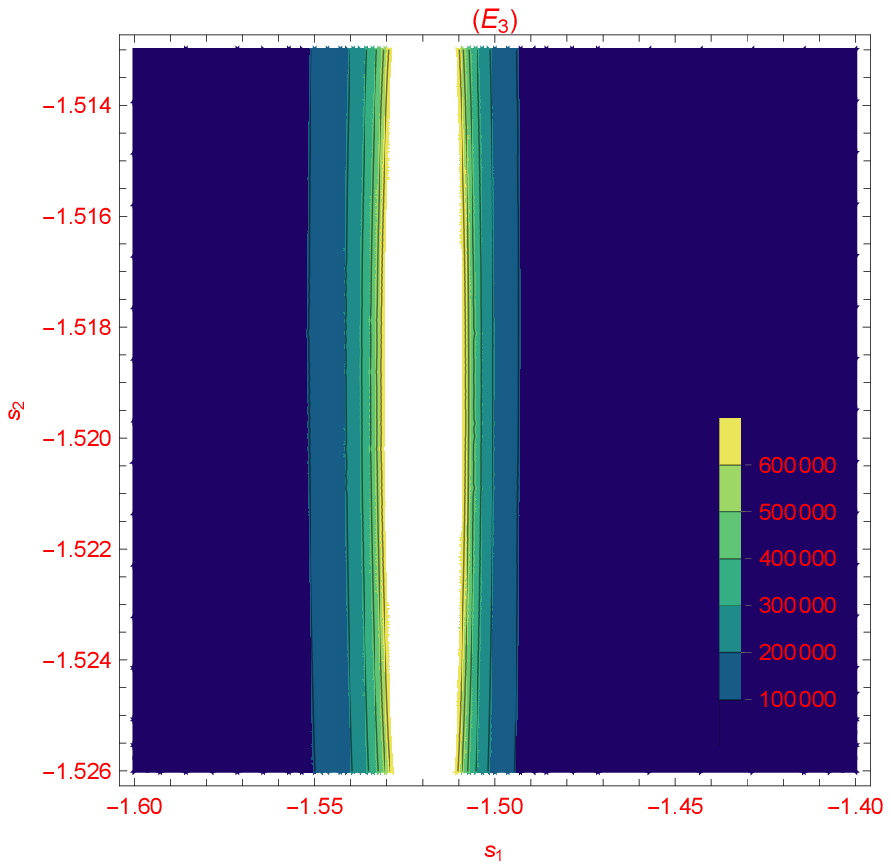}\label{fig33}}
	\caption{The check of fermion in the Penrose collision.}
	\label{fig5}
\end{figure}

\newpage
\begin{table}[tbp]
\centering
\begin{tabular}{|c|c|c|c|c|}
\hline
      &check	         &$s_{3=1}\textbf{=}s_{4=2}$   &${(s_{3=1})}_{max}$          &${(s_{4=2})}_{max}$           \\
\hline
\hline
$6.$  &$s=\frac{1}{2}$   &$0.4\leq s_{3=4}\leq0.6$     &$-1.526\leq s_3\leq-1.519$   &$0.4\leq s_3\leq0.6$          \\
      &                  &                             &$0.4\leq s_4\leq0.6$         &$-1.526\leq s_4\leq-1.513$    \\
\hline
      &$E_3$             &$0.1 m_3$ $(a)$              &$0.94 m_3$   $(b)$           &$4.85 m_3$   $(c)$            \\
\hline
\hline
$7.$  &$s=-\frac{1}{2}$  &$-0.6\leq s_{3=4}\leq-0.4$   &$-1.526\leq s_3\leq-1.519$   &$-0.6\leq s_3\leq-0.4$        \\
      &                  &                             &$-0.6\leq s_4\leq-0.4$       &$-1.526\leq s_4\leq-1.513$    \\
\hline
      &$E_3$             &$40 m_3$ $(d)$               &$14 m_3$  $(e)$              &$3.45 m_3$  $(f)$             \\
\hline
\hline
$8.$  &$s=+\frac{3}{2}$  &$1.4\leq s_{3=4} \leq1.6$    &$-1.526\leq s_3\leq-1.519$   &$1.4\leq s_3\leq1.6$          \\
      &                  &                             &$1.4\leq s_4\leq1.6$         &$-1.526\leq s_4\leq-1.513$    \\
\hline
      &$E_3$             &$-1.2 m_3$   $(g)$           &$1.05 m_3$ $(h)$             &$13 m_3$   $(i)$              \\
\hline
\hline
$9.$  &$s=-\frac{3}{2}$  &$1.4\leq s_{3=4} \leq1.6$    &$-1.526\leq s_3\leq-1.519$    &$1.4\leq s_3\leq1.6$         \\
	  &   	             &                             &$1.4\leq s_4(=s_2)\leq1.6$    &$-1.526\leq s_4\leq-1.513$   \\
\hline
      &$E_3$             &$8\times 10^{5} m_3$  $(j)$  &$6\times 10^{5} m_3$  $(k)$   &$6\times 10^{5} m_3$  $(l)$  \\
\hline
\end{tabular}
\caption{\label{tab2}Check of fermion in the Penrose collision.}
\end{table} 
Now it is the turn of fermion particles. The description of our studies according to Fig.\ref{fig5} and table\ref{tab2} is as follows:\\
\textbf{6.} The first fermion studied is $s=\frac{1}{2}$. For the case where both particles have equal spins (a), $E_3=0.1 m_3$. If the extraction particle is $s_{max}$ (b), $E_s=0.94 m_3$. Furthermore, if $s_{max}$ falls into the black hole (c), we have $E_s=4.85 m_3$. \\
\textbf{7.} In spin mode $s=-\frac{1}{2}$, we have $E_3=40 m_3$, for equal to spin mode (d). For the subsequent two cases (e and f), we have $E_3=14 m_3$ and $E_3=3.45 m_3$.\\
\textbf{8.} In $s=+\frac{3}{2}$, we obtained the energy of extracted particles $E_3=-1.2 m_3$ (g), $E_3= 1.05m_3$ (h) and 6$E_3=13 m_3$ (i), respectively, that Mode (g) is unacceptable.\\
\textbf{9.} The last spin we consider is $s_{max}=-\frac{3}{2}$, which gives the highest extraction energy. So we have energy in different modes $E_3=8\times 10^{5} m_3$  (j), $E_3=6\times 10^{5}m_3$ (k), and $E_3=6\times 10^{5}m_3$ (l), respectively.\\
Our calculations show that energy extraction is possible in approximately all spin modes.
\section{Concluding remark}\label{V}
We studied the collision of two spinning-massive particles near Kerr-Newman black holes with WGC conditions. We use the near-extremal condition for the Kerr-Newman black hole to avoid the naked singularity $M^2=\lim_{\varepsilon \longrightarrow 0} (a^2+Q^2+\varepsilon)$. We investigate the collision Penrose process in the ergosphere region of the black hole, where two particles fall from infinity and scatter to the other two. One of the new scattered particles with negative energy falls into the black hole during the collision, and the other escapes to infinity. We first comprehensively describe the BSW mechanism for the general ansatz of a stationary axisymmetric black hole. We show that the energy, $4$-momentum, spins particles, and angular momentum is conserved in this collision. Furthermore, our study shows that particles' energy and angular momentum exert constraints on their orbits. There is a critical value for the energy and angular momentum of particles $\Gamma_c=J_c/E_c$; for the orbit of particles to reach the horizon, it should be $\Gamma \leq\Gamma_c$. However, the energy maximum is when 1th particle is critical $(\Gamma=\Gamma_c)$, 2th particle is non-critical $(\Gamma=\Gamma_c (1+\xi))$ and 3th particle is near-critical $(\Gamma=\Gamma_c (\sigma\epsilon+\gamma\epsilon^2+...))$. Finally, we calculate the energy extracted from the third particle for all terms $\mu=(\sigma\epsilon+\gamma\epsilon^2+...)$ and we show that for values $0\leq\mu\leq0.1$ and $\varepsilon=0.1$, the energy extracted is $E_3=1800000m_3$ for spins $s_1(=s_3)=-1.526,-1.519$ and $s_2(=s_4)=-1.513,-1.526$. We also examined the energy extracted from the black hole for different spins. We showed that almost any particle could extract energy from the black hole by colliding near it. This conclusion helps us study the extremality black holes, astrophysics black holes, and WGC conditions.

\appendix
\section*{Appendix}
\section{The extractive energy:}\label{App}
The energy of the third particle is in energy terms of the primary particles and spin that escapes to infinity from the Kerr-Newman black holes.
\begin{equation*}\label{e.30}
E_3=\frac{-A\pm \sqrt{A^2-4BC}}{2B}
\end{equation*}
\begin{equation*}\label{e.31}
	\begin{split}
	  A=&-1.13s_1^2+(1+\mu)\textbf{[}\chi_{s_1}(9.59+8.24s_1)-16.78-28.9s_1-8.54s_1^2+3.35s_1^3-0.22s_1^4+(16.78+28.9s_2\\
		&+8.54s_2^2-3.35s_2^3+0.22s_2^4-(6.8+10.57s_2-0.09s_2^2-3.68s_2^3)E_1-(1+\xi)s_2E_2(10.6-0.09s_2+0.1s_2^3\\
		&-3.68s_2^2)-(1+E_1)6.8E_2\xi-\chi_{s_2}(9.59+8.24s_2-1.13s_2^2-(E_1+E_2(1+\xi))(7+6.04s_2-0.8s_2^2)))\\
		&(0.62 +0.56s_1+0.1s_1^2)^2(1.46-0.32s_1^2)^2\textbf{]}\\
	  B=&(0.62+0.56s_2+0.1s_2^2)^2(1.46-0.32s_2^2)^2\textbf{[}(3.4+5.28s_1-0.04s_1^2-1.84s_1^3+0.05s_1^4)(1+\mu^2)+(10.57s_1\\
		&+6.8-0.09s_1^2-3.68s_1^3+0.1s_1^4)\mu-((3.5+3.02s_1-0.4s_1^2)(1+\mu^2)+(7+6.04s_1-0.8s_1^2)\mu)\chi_{s_1}\textbf{]}+\\
		&(0.62+0.56s_1+0.1s_1^2)^2(1.46-0.32s_1^2)^2\textbf{[}(3.4+5.28s_2-0.04s_2^2-1.84s_2^3+0.05s_2^4)(1+\mu^2)+(6.8+\\
		&10.57s_2-0.09s_2^2-3.68s_2^3+0.1s_2^4)\mu-((3.5+3.02s_2-0.4s_2^2)(1+\mu^2)+(7+6.04s_2-0.8s_2^2)\mu)\chi_{s_2}\textbf{]}\\
	  C=&(\delta_1F_1-\delta_2F_2)^2-G_{s_2}H_{s_1}-G_{s_1}(H_{s_2}-K_{E_1}-K_{E_2}+F_3)\\
        &G=-(0.62+0.56s+0.1s^2)^2(1.46-0.32s^2)^2\\
        &H=(20.01+38.8s+23.22s^2+4.49s^3+0.19s^4-0.005s^5-0.0009s^6)\\
        &K=E(16.8+28.9s_2+8.54s_2^2-3.35s_2^3+0.22s_2^4-E(3.4+5.28s_2-0.04s_2^2-1.84s_2^3+0.05s_2^4))\\
        &\chi=\sqrt{0.34-1.63s^2-1.24s^3+0.01s^4}\\
        &\mu=(\sigma\epsilon+\gamma\epsilon^2+...)\\
        &F_1=(H_{s_1}-K_{E_1}+E_1(9.59+8.24s_1-1.13s_1^2-E_1(3.5+3.02s_1-0.4s_1^2)))\chi_{s_1}\\
        &F_2=(K_{E_2}-(16.78+28.9s_2+8.54s_2^2-3.35s_2^3+0.22s_2^4-(6.8+10.57s_2-0.09s_2^2-3.68s_2^3+0.1s_2^4)E_2)\\
        &E_2(1+\xi)+(3.4+5.28s_2-0.04s_2^2-1.84s_2^3+0.05s_2^4)E_2^2\xi^2+E_2((9.59+8.24s_2-1.13s_2^2)(1+\xi)-(7E_2\\ 
        &+6.04s_2E_2-0.8s_2^2E_2)\xi-(3.5+3.02s_2-0.4s_2^2)E_2(1+\xi^2)))\chi_{s_2}\\
        &F_3=(6.8+10.57s_2-0.09s_2^2-3.68s_2^3+0.1s_2^4)E_1E_2+\chi_{s_2}(E_1^2+E_2^2)+E_2^2)(-(3.5+3.02s_2)+0.4s_2^2E_2^2+\\
        &(9.59+8.24s_2-1.13s_2^2)(E_2+E_1)-E_1E_2(7+6.04s_2-0.8s_2^2)+\xi(((E_1+E_2)(6.8+10.57s_2-0.09s_2^2-\\
        &3.68s_2^3+0.1s_2^4)-16.78-28.9E_2-8.54s_2^2+3.35s_2^3-0.22s_2^4)E_2+\chi_{s_2}((9.59+8.24s_2-1.13s_2^2)E_2-(7\\
        &+6.04s_2-0.8s_2^2)))(1+E_2^2)+E_2^2\xi^2(3.4+5.28s_2-0.04s_2^2-1.84s_2^3+0.05s_2^4+\chi_{s_2}(3.5-3.02s_2+0.4s_2^2)))
    \end{split}
\end{equation*}

\end{document}